\documentclass[fleqn,11pt]{article}
\usepackage[utf8]{inputenc}
\usepackage[T1]{fontenc}
\usepackage{amsthm}

\usepackage[a4paper]{geometry}
\usepackage{authblk}
\usepackage[round]{natbib} 
\usepackage{comment}

\usepackage{amsmath,amssymb} 
\usepackage{graphicx}
\usepackage[colorlinks=true, allcolors=blue]{hyperref}
\usepackage{csquotes}

\begin{document}

\title{Modeling adaptive forward-looking behavior in epidemics on networks  }


\author[1]{Lorenzo Amir Nemati Fard}
\author[2]{Alberto Bisin}
\author[3,4 $\ast$]{Michele Starnini}
\author[5,$\dagger$]{Michele Tizzoni}

\affil[1]{Department of Physics, University of Pisa, Largo Bruno Pontecorvo 3, 56127, Pisa, Italy}
\affil[2]{Department of Economics, NYU, NBER, and CEPR, New York University, 19 West Fourth Street,
New York, NY 10012, USA}
\affil[3]{Department of Engineering, Universitat Pompeu Fabra, 08018,
Barcelona, Spain}
\affil[4]{CENTAI Institute, Turin, Italy}
\affil[5]{Department of Sociology and Social Research, University of Trento, Trento, Italy}

\affil[$\ast$]{\small To whom correspondence should be addressed: \href{email:michele.starnini@upf.edu}{michele.starnini@upf.edu}\\}
\affil[$\dagger$]{\small To whom correspondence should be addressed: \href{michele.tizzoni@unitn.it}{michele.tizzoni@unitn.it}}

\maketitle

\begin{abstract}
Incorporating decision-making dynamics during an outbreak poses a challenge for epidemiology, faced by several modeling approaches siloed by different disciplines. 
We propose an epi-economic model where {high-frequency} choices of individuals respond to the infection dynamics over heterogeneous networks.
{Maintaining a rational forward-looking component to individual choices,  agents follow a behavioral rule-of-thumb in the face of limited perceived forecasting precision in a highly uncertain epidemic environment.} We describe the resulting equilibrium behavior of the epidemic by analytical expressions depending on the epidemic conditions. We study existence and welfare of equilibrium, identifying a fundamental negative externality. We also sign analytically the effects of the behavioral rule-of-thumb  at different phases of the epidemic and characterize some comparative statics.
Through numerical simulations, we contrast different information structures: global awareness -- where individuals only know the prevalence of the disease in the population -- with local awareness, where individuals know the prevalence in their neighborhood. 
We show that agents' behavioral response through forward-looking choice can flatten the epidemic curve, but local awareness, by triggering highly heterogeneous behavioral responses, more effectively curbs the disease compared to global awareness.
\end{abstract}

\section{Introduction}
\label{sec: introduction}

Agents' behavioral adaptation in response to infectious disease outbreaks is one of the key factors that shape the course of an epidemic \citep{ferguson2007capturing}. 
Individual choices regarding the adoption of self-protective measures, such as wearing masks, avoiding close social contacts, or vaccinating, contribute to reducing disease transmission in the population. 
In turn, such choices vary over time as the epidemic progresses in that they depend on the perceived severity of the epidemic, the available information about it, and individual risk perception.
As the epidemic fades out, behavioral responses may relax, with the consequence of leading to disease resurgence. 
Self-initiated behavioral responses have been observed across all kinds of epidemics, from small-scale outbreaks involving few individuals, such as the 2015 Middle East respiratory syndrome outbreak in South Korea \citep{kim2017exposure}, to worldwide pandemics, such as the 1918 pandemic \citep{crosby2003america}, the 2009 A/H1N1 pandemic \citep{jones2009early}.
During the COVID-19 pandemic, in particular, several papers documented how fear of contagion has affected individual mobility \citep{aum2020covid, barrios, bartik2020small, 
coibion2020cost, goolsbee2020fear, gupta2020effects, NBERw27061, 
rojas2020cure}.

Incorporating the dynamics of individual decision-making during an outbreak represents a key challenge of epidemic modeling \citep{funk2015nine}.
To this aim, a variety of mathematical epidemic models have been proposed that capture the effects of behavioral changes \citep{funk2010modelling,verelst2016behavioural,manfredi2013modeling}. 
Generally speaking, these models are well-established in the literature and can be classified into three broad classes.
In the simplest case, classical compartmental models have been expanded to consider additional behavioral classes in the population, characterized by different behavioral responses to disease prevalence, whose transmission parameters do not change over time \citep{perra2011towards, d2022individual}.
The second class of models includes those that aim to capture the interaction between individual adaptation and individual knowledge of the disease, often represented as two coupled dynamical processes \citep{granell2013dynamical,epstein2008coupled}.\footnote{Information about the disease can be local or global, and sometimes it is assumed to spread through the population \citep{funk2009spread}.} 
Finally, a distinct class of models developed by economists and epidemiologists aims to explicitly describe the individual decision-making process assuming some form of forward-looking rationality \citep{aadland2013syphilis,fenichel2011adaptive, geoffard1996rational,goenka2012infectious, poletti2012risk,rowthorn2012optimal}.
These ``epi-economic'' models simulate the process by which people choose the best course of action by adjusting to the current epidemic state and seeking the best possible future outcome.

The COVID-19 pandemic reopened this debate on how human behavior should be included in epidemic models, with researchers with many different backgrounds contributing to the modeling effort.
On the one hand, the literature expanded on agent-based models with additional behavioral classes requiring an {\em ad hoc} behavioral response \citep{Azzimontietal2020, bisin-moro-lockdown-lucas-critique-2022, keppo2020behavioral}.
On the other hand, a new generation of forward-looking rational expectation epi-economic models has been developed \citep{acemoglu2020multi, aguirre, alfaro2020social, alvarez2021simple, bethune2020covid, brotherhood2020economic, farboodi2021internal, phelan2022optimal,toxvaerd2020equilibrium}; see \citep{mcadams2021blossoming} for a survey. 
{In these forward-looking models individual expectations about the dynamics of the epidemic are generally assumed rational and precise - often deterministic in fact. This suggests an environment where individual choices are taken at relatively low frequencies and forecasting errors about, e.g.,  the transmission and recovery rates, are smoothed over time. 
Furthermore, these models generally rely on the homogeneous mixing hypothesis, under which all susceptible individuals have an equal chance of contracting the infection.\footnote{Notably, \cite{alfaro2020social} extends the analysis to allow for differential contact rates by demographic group.}} 

{In this work, we aim instead at studying high-frequency choices of individuals responding to the infection dynamics over heterogeneous networks, capturing heterogeneous behavioral responses in the population \citep{
meyers2005network}. To this end, we construct an epi-economic model where individual choices, while maintaining
a rational forward-looking component, follow a behavioral rule-of-thumb in the face of low perceived forecasting precision in a highly uncertain epidemic environment, e.g., the initial phase of an epidemic.
Furthermore, we allow different individuals to have varying chances of contracting the disease as their choices respond to the dynamics of the infection on a heterogeneous networked substrate.}

More specifically, individuals' choices are forward-looking, in that they optimally choose their social activity to maximize trading off the risk of being infected in the future with their preference for maintaining the highest possible social activity.  
{However, it is assumed that the individuals facing this decision follow a myopic rule-of-thumb: they anticipate constant epidemic conditions so that the optimization problem they face at time $t$ replicates identically in the future. These epidemic conditions are rationally obtained at equilibrium by postulating that all other agents solve their choice problem under the same conditions.\footnote{\cite{gonzalez2023optimal} also studies high-frequency individual choices in the context of an epidemic, but under the assumption that such choices are static.} Measuring time in days with a high discount rate - whereby utility is halved, e.g., in about a week - this rule-of-thumb can be interpreted as a sort of short-run high-frequency forecasting strategy: large changes in the epidemic environment in the short-run have small probability and, in any case, any changes will be ex-post observed and reacted upon in a matter of days.  
In this context, the discount rate $\delta$ can be interpreted as an ex-ante measure of the agent's perceived precision of the constant epidemic assumption, rather than a measure of psychological preference for the present.\footnote{An alternative approach to modeling limited forecasting precision would consist of fixing a finite planning horizon, i.e., choosing the number of future days that are taken into account when planning \citep{nardin2016planning, fenichel2011adaptive}. However, this choice has the
drawback of adding another parameter to the model (the planning horizon), making it more difficult to fit with empirical data.} 
} 

Under this behavioral assumption, we can describe the resulting equilibrium behavior of the epidemic by an analytical expression depending on the epidemic conditions (prevalence and disease parameters). We then study existence and welfare, identifying a fundamental negative externality in individuals' choices about social activity. We also sign analytically the effects of myopic behavior at different phases of the epidemic and characterize some comparative statics results.\footnote{Relatedly, and under a similar assumption, \cite{acemoglu2020testing} study the effects of testing in a SIR model which allows for endogenous network formation.} 
We can then explicitly contrast different information structures in the population. We differentiate between \emph{global awareness}, in which individuals only have a bird-eye view of the epidemic unfolding throughout the population, and \emph{local awareness}, in which individuals explicitly know how many of their contacts are infected. 

We show that agents' behavioral response through forward-looking choice can flatten the epidemic curve by lowering the peak prevalence, thus potentially reducing the load on the health system at the epidemic peak. Most importantly, we show that local awareness triggers highly heterogeneous behavioral responses. 
Nonetheless, the aggregate composition of such heterogeneous local responses induces a more effective curbing of the epidemic in its early phase with respect to behavior under global awareness. Indeed, the stronger effect of local awareness can be attributed to the heterogeneous behavioral response: individuals in direct contact with local outbreaks reduce their social activity, hindering early disease propagation. 
Therefore, even when the prevalence is low in the population and is localized in a few individuals, the behavioral response of the neighbors of the infected individuals is sufficient to curb the spread of the disease.
In contrast, within the global setting, the prevalence has to grow sufficiently in the population to trigger a behavioral response from individuals.

\section{The model}
\label{sec:model}
We propose an analytically tractable epidemic model linking the social activity of individuals and the spreading of the epidemic. Individuals choose their social activity according to the prevalence of the disease. Susceptible individuals trade off the risk of being infected in the future with their preference for social activity. We primarily study a Susceptible-Infected-Recovered (SIR) model \citep{kermack1927contribution}, 
---the standard workhorse in the literature--- 
but in Section \ref{sec:sis} we show that alternative infection dynamics, such as a Susceptible-Infected-Susceptible (SIS) model, have similar implications. 
The model is designed to study a highly uncertain epidemic environment where individuals face low perceived forecasting precision of epidemic conditions and hence, effectively, a very short-horizon decision problem. 

We consider a population of $N$ individuals, indexed by  $j=1, \ldots N$. Each individual belongs to one of three compartments: susceptible ($s$), infected ($i$), or recovered ($r$). 
A $s$ individual has a probability of becoming infected upon contact with a $i$ individual in one time step, which we assume equals one day.
$i$ individuals recover with probability $\mu$ in the same time interval. $r$ individuals can not be infected anymore. 
Individual behavior affects the probability of infection. Let $a^j_t$ denote the \textit{social activity} of an individual $j$ at time $t$, representing this individual's propensity to engage in social interactions with peers.
For any agent $j$, social activity is bounded ${0\le a^j_t\le1}$, with ${a^j_t=1}$ corresponding to normal behavior in the absence of the disease, and ${a_t=0}$ corresponding to a situation in which disease transmission is not possible, equivalent to quarantine. 

\begin{figure}[tbp]
	\centering
	\includegraphics[width=0.85\textwidth]{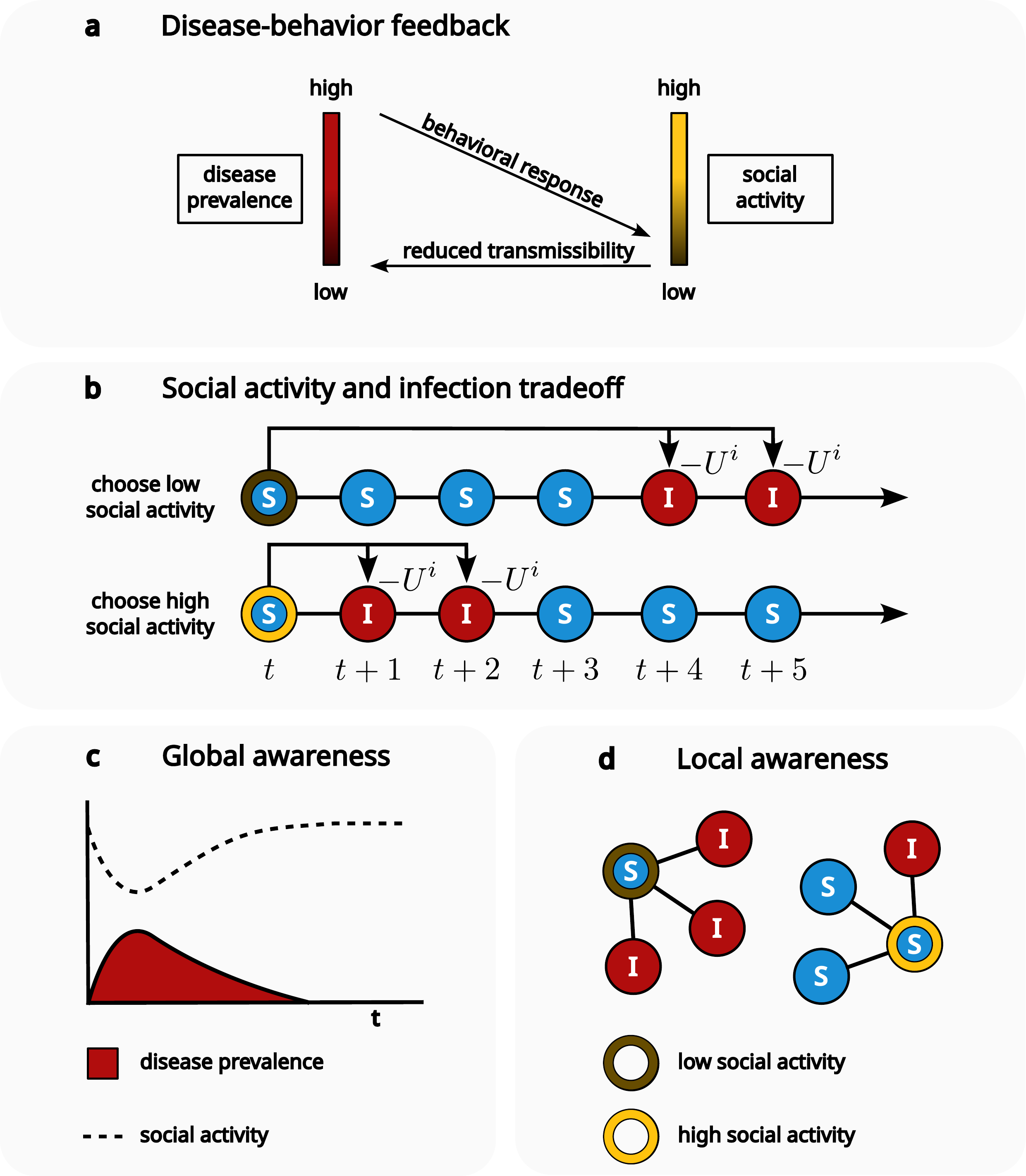}
	\caption{
  Schematic illustration of the model. \textbf{(a)} Feedback loop involving epidemic spreading and {social activity}. If the prevalence is high, individuals can decrease their social activity, which in turn reduces the transmissibility of the disease, decreasing prevalence. With low prevalence, individuals can choose high levels of social activity, increasing the prevalence. 
  \textbf{(b)} Individuals choose their social activity to maximize their future expected utility. They can choose a low value of social activity to delay the expected infection (and the expected penalty $U^i$), or benefit from high social activity at a higher infection risk.  
  Awareness and behavioral response can be of two kinds. \textit{Global awareness} \textbf{(c)}: individuals know the prevalence of the disease in the population but do not know who is infected, so social activity is homogeneous across the population. 
  \textit{Local awareness} \textbf{(d)}: individuals know how many of their contacts are infected, resulting in different values of social activity for each individual.}
\label{fig: explicative_fig}
\end{figure}

We assume that disease transmission between individuals depends on their social activity. 
Let $\textbf{a}_t$ denote the entire configuration of actions in the population at time $t$ (a $N$-dimensional vector) and $\textbf{i}_t$ the distribution of infections at time $t$ (a $N$-dimensional vector with $1,0$ for infected/non-infected individuals). 
At this level of abstraction, the infection probability at time $t+1$ of a susceptible individual $j$ 
will depend on the actions of all the agents in the society $\textbf{a}_t$, and on the distribution of infections in the population $\textbf{i}_t$: 
$P^j_{t+1}(\textbf{a}_t, \textbf{i}_t)$. For any $j$ and at any time $t$, we assume that $P^j_{t+1}(\textbf{a}_t, \textbf{i}_t)$ is continuously differentiable in $\textbf{a}_t$, for any given $\textbf{i}_t$.
Each individual $j$ chooses optimally to limit social activity $a^j_t$ to reduce the probability of being infected: hen the prevalence is high, they will adopt prudent behavior. 
The feedback loop between the disease spread and
social activity is illustrated in Figure \ref{fig: explicative_fig} (a).

At each time $t$, each individual $j$'s \textit{utility} is the sum of two terms $U(a^j_t)$ and $U^h_t$: the first depends only on social activity and the second depends only on the health status ${h = \{s,i,r\}}$. 
Utility from social activity $U(a^j_t)$ is logarithmic with a linear cost of activity: 
\begin{equation}\label{eq: U_a}
	U(a^j_t) = \log(a^j_t) - a_t +1. 
\end{equation}
Since $0 \leq a^j_t \leq 1$, $U(a^j_t)$ is normalized to represent the utility cost of reducing social activity from its normal status ($=1$) to $a^j_t$. 

The state-dependent term $U^h_t$ of the utility function corresponds to a fixed penalty of magnitude $U^i$ for each time step (day) in the infected state and to be equal to zero otherwise:
\begin{equation}
\label{eq: U^h}
U^h_t = \begin{cases}
	-U^i & \text{if } h = i \text{ at time $t$}\\
	0 & \text{otherwise}.
\end{cases}
\end{equation} 
Given $U_\tau^h$, we calculate the average cost of infection, which we indicate by $\alpha$.
An individual who has just been infected at time $t$ gets a penalty of magnitude $U^i$ for each time step spent in the infected state, as for Eq. \eqref{eq: U^h}. 
In each time step, the infected individual recovers from the disease with probability $\mu$: the probability of still being infected at time $\tau$ is $(1 - \mu)^{\tau - t}$. Including discounting, the average loss of utility caused by an infection is:
\begin{equation}\label{eq: alpha}
  \alpha \equiv U^i \sum_{\tau =t}^\infty [\delta(1 - \mu)]^{\tau - t}
	= \frac{U^i}{1 - \delta(1-\mu)}.
\end{equation}
The average cost of infection 
is thus proportional to the penalty $U^i$ and it becomes smaller for increasing discount factor $\delta$ and recovery rate $\mu$. 

The optimal value of social activity for the individual $j$ at each time $t$ is then given by the maximum with respect to $a^j_t$ of the following objective function:
\begin{equation}
\label{eq: objective}
	J^j_t = \mathbb{E} \sum_{\tau=t}^\infty\ \delta^{\tau-t} \left[ U(a^j_{\tau}) + U_{\tau}^h \right],
\end{equation}
where $0<\delta<1$ is the discount rate. 
 \newpage
\noindent The objective function in Eq. \eqref{eq: objective} represents the expected value, over all possible future health states, of the present discount utility. \bigskip 

\noindent {\bf Assumptions: Behavior.} {\em Each individual $j$ at any time $t$,
\begin{itemize}
\item[i)] does not observe his/her own health status prior to choosing social activity;
\item[ii)] myopically anticipates his/her choice  problem to replicate identically in the future, from $t+1$ to $\infty$;
\end{itemize}}

\noindent Assumption (i) is particularly suitable for studying asymptomatic infections.\footnote{We follow \cite{farboodi2021internal} in this respect. 
Note that for $r$ individuals this is irrelevant since they do not participate in active links.}
It implies that individuals estimate the probability of being susceptible to the disease, as well as the probability of becoming infected in the future, under different information scenarios. 
Leaving out the possibility of multiple infections, the probability of still being susceptible when optimizing is denoted $\sigma^j_t (\textbf{s}_t)$, with $\textbf{s}_t$ representing the distribution of susceptible individuals (a $N$-dimensional vector 1, 0 for susceptible/non-susceptible individuals). 
Assumption (ii) is motivated by the low perceived forecast precision of epidemic conditions. It implies that individuals follow a myopic behavioral rule-of-thumb, to avoid forecasting the infection dynamics.\footnote{These assumptions are further discussed in Section \ref{sec: behavioral_parameters} (Model calibration).}  
Under these assumptions, we can characterize the choice of social activity of each agent $j$. \bigskip 

\noindent {\bf Proposition: Social activity.} {\em Social activity is determined as the maximum with respect to $a^j_t$ of the objective function in Eq. \eqref{eq: objective}.
In particular, if the behavioral component of the utility function is given by Eq. \eqref{eq: U_a}, then the optimal value of social activity at time $t$ is determined by:
\begin{equation}\label{eq: social_activity_general}
	a^j_t = \frac{1}{1+ \alpha \delta \sigma^j_t(\textbf{s}_t) \frac{\partial P^j_{t+1}}{\partial a^j_t} (\textbf{a}_t, \textbf{i}_t)}, \; \; \forall \; j=1, \ldots, N.
\end{equation} } 

\begin{proof} We derive next the optimal policy (\ref{eq: social_activity_general}). We first construct a closed-form expression for the objective function in Eq. \eqref{eq: objective}.  
Since we assumed individuals to believe current conditions to persist in the future, the term involving $U(a^j_\tau)$ depends neither on time nor on the health status (which we assumed  un-observable),  
so it is just a geometric series:
\begin{equation}
\label{eq: objective_U_a}
	\mathbb{E}\sum_{\tau = t}^{\infty} \delta^{\tau - t} U(a^j_\tau) = \frac{1}{1 - \delta}U(a^j_t).
\end{equation}

Let us now consider an individual that is evaluating the expected future utility loss because of the risk of infection at time $t$.
If a susceptible individual gets infected at time $\tau$, the expected penalty (given by Eq. \eqref{eq: alpha}) gets discounted by a factor $\delta^{\tau - t}$. 
The remaining term to close the calculation is the probability of becoming infected at time $\tau$.
The probability $P^j_{t+1}$ of becoming infected in one time step at time $t$ (the time when the expected utility is evaluated) is assumed to remain constant in the future, since individuals believe current conditions to persist in the future. Therefore individuals estimate their probability of becoming infected at time $\tau$ to be $P^j_{t+1}(1-P^j_{t+1})^{\tau - t -1}$ (individuals must not have already been infected until time $\tau - 1$).
Therefore, at time $t$, the expected utility loss due to the risk of infection at time $\tau$ is the product the average loss for infection $\alpha$ and the probability of getting infected at time $\tau$, and thus discounted by $\delta^{\tau - t}$. Since this penalty is only relevant for susceptible individuals,\footnote{The expected penalty for individuals that are $r$ at time $t$ is zero, while the one for $i$ ones (given by the cost of infection $\alpha$ multiplied by the probability of being infected at time $t$) is irrelevant since it does not depend on social action.} we multiply it by the probability of being susceptible $\sigma^j_t$: 
\begin{equation}
	\mathbb{E}\sum_{\tau = t}^{\infty} \delta^{\tau - t} U^h_\tau =%
	- \sigma^j_t \delta P^j_{t+1} \sum_{\tau = t +1}^{\infty} [\delta(1-P^j_{t+1})]^{\tau - t -1} \alpha. \label{EULoss} 
\end{equation}
By solving the geometric series and assuming that the probability of getting infected at time $t$ is small, $P^j_{t+1}\ll1$, one obtains\footnote{This assumption, typically valid for respiratory infections such as influenza, can be expanded to more transmissible diseases by using a smaller time unit (e.g., few hours instead of a day).}:
\begin{equation}
\label{eq: objective_U_H}
	\mathbb{E}\sum_{\tau =t}^{\infty} \delta^{\tau - t} U^h_\tau =
	-\frac{\alpha \delta \sigma^j_t P^j_{t+1}}{1 - \delta} .
\end{equation}
Combining Eq. \eqref{eq: objective_U_a} and Eq. \eqref{eq: objective_U_H} into Eq. \eqref{eq: objective}, the closed form expression for the objective function at time $t$ is obtained:
\begin{equation}\label{eq: objective_simplified}
	J^j_t = \frac{1}{1-\delta}\left[U(a^j_t) - \alpha \delta \sigma^j_t(\textbf{s}_t) P^j_{t+1} (\textbf{a}_t, \textbf{i}_t)\right].
\end{equation} 
Equation (\ref{eq: social_activity_general}) now follows directly as the first order condition of the (convex) maximization problem of equation (\ref{eq: objective_simplified}) with respect to $a^j_t$.
 \end{proof} 

Keeping the contact process fixed, the functions $\sigma^j_t(\textbf{s}_t)$  and $P^j_{t+1}(\textbf{a}_t, \textbf{i}_t)$ in equation (\ref{eq: social_activity_general}) depend on the specifics of the network structure of individuals and their information about the disease, which we shall introduce in Section \ref{sec:awareness}. The terms $\textbf{s}_t, \: \textbf{i}_t$, in turn, are determined at equilibrium. \bigskip 

\noindent {\bf Theorem: Existence.} {\em With a finite number of agents $N$, an equilibrium social activity vector $\textbf{a}_t$ exists.}  
\begin{proof} Under our assumptions, the contact process induces a continuous dependence of $\frac{\partial P^j_{t+1}}{\partial a^j_t}
(\textbf{a}_t, \textbf{i}_t)$ on $a^j_t$.
For each individual, social activity has a maximum $a^j_{max}=1$ and a minimum $a^j_{min}$, nothwistanding log utility (Eq. \eqref{eq: U_a}). The existence of the minimum stems from the continuity of Eq. \eqref{eq: social_activity_general} on the compact set $\{(s_t,i_t,a_t) \mid s_t \in [0,1],i_t \in [0,1],a_t \in [0,1]\}$.
Stacking equations (\ref{eq: social_activity_general}) for $j=1, \ldots, N$, we obtain a system from $\textbf{a}_t \in
\{(a^1_t,…,a^N_t) \mid a^j_t \in [a^j_{min},1] \, \forall \, j \in \{1,…,N\}\}$ into itself which satisfies the conditions of Brouwer's Theorem for the existence of a fixed point representing the equilibrium. 
\end{proof}  

We now specify how the disease transmission is mediated through social interactions, that is, through a contact network. 
A susceptible individual $s$ with social activity $a^s_t$ coming into contact at time $t$ on the contact network with an infected individual $i$ with social activity $a_t^i$, is infected at time $t+1$ with probability $\beta a^s_t a_{t}^i.$ On the other hand, the number of contacts of individuals depends on their network structure, which is exogenously given in the model.\footnote{We note that the social activity $a^j_t$ can have multiple interpretations as long as they result in a decreased probability of disease transmission. 
For instance, a lower social activity could represent less frequent contact with other individuals or it could represent the adoption of prophylactic measures, such as the use of face masks in the case of airborne disease \citep{Tirupathi2020comprehensive}, that limit the probability of infection on contact without actually reducing the contact rate.} 
This contact process directly induces a probability of becoming infected next period $P^j_{t+1} (\textbf{a}_t, \textbf{i}_t)$ which is linear in $a^j_t$ - so that $\frac{\partial P^j_{t+1}}{\partial a^j_t}
(\textbf{a}_t, \textbf{i}_t)$ is a non-negative constant depending on the network structure, the distribution of the infection (the vectors $\textbf{s}_t, \textbf{i}_t$) and of social activity $\textbf{a}_t$.  
The following comparative statics is then a straightforward consequence of Eq. (\ref{eq: social_activity_general}). \bigskip 

\noindent {\bf Proposition: Comparative statics.} {\em Each individual $j$ will adopt a higher optimal value of social activity $a^j_t$ at each time $t$, when the constant parameters $\alpha$ and $\delta$ and the probability of being currently susceptible $\sigma^j_t$ are lower. 
Furthermore, $a^j_t$ will be higher when the network structure of contacts induces a lower probability of being infected given the social activity of $j$ (when the constant $\frac{\partial P^j_{t+1}}{\partial a^j_t}
(\textbf{a}_t, \textbf{i}_t)$  is lower). } \bigskip 

To pursue a welfare analysis of equilibrium, it is natural to consider a constrained Pareto efficiency notion, whereby the planner is constrained to the same myopic expectations as individuals. In this case, constrained efficiency is defined as the choices  $a^j_t$, for all $j$, of a planner who  maximizes $$ \sum_{j} J^j_t = \sum_j \: \mathbb{E} \sum_{\tau=t}^\infty\ \delta^{\tau-t} \left[ U(a^j_{\tau}) + U_{\tau}^h \right],$$ under myopic expectations as in Assumption ii). Under these conditions we identify a fundamental negative externality in individuals’ choices, whereby they do not internalize the effects of their social activity on the probability of infection of other individuals in their contact network.\bigskip 

\noindent {\bf Theorem: Constrained inefficiency.} {\em Equilibrium social activity is constrained inefficient - in fact, inefficiently too high for any agent $j$ such that $\frac{\partial P^j_{t+1}}{\partial a^{j}_t}
(\textbf{a}_t, \textbf{i}_t) > 0$.   } 
 \begin{proof} The result follows directly by noticing that the  contact process induces a 
  $\frac{\partial P^j_{t+1}}{\partial a^{j'}_t}
(\textbf{a}_t, \textbf{i}_t) \geq 0$, for any $j \not = j'$ --- in fact,  strictly so for any  $j, j'$ in contact.
  \end{proof}

We can also sign the effect of the myopic rule-of-thumb adopted by each individual: it will tend to induce higher social activity in equilibrium at the outset of the epidemic - when the probability of getting infected is increasing -  and lower social activity after the peak of the epidemic. More precisely, 
\bigskip 

\noindent{\bf Proposition: Myopic vs. rational expectations.} {\em Suppose that each individual $j$ at time $t$ rationally anticipates the dynamics of the epidemic, that is, all $P^j_{t+\tau}$,  $\tau \geq 1$. Suppose furthermore that $P^j_{t+\tau}$ has a single peak at $t+\tau^*$. Then, there exists a $\tau^*$ large enough and a discount $\delta$ low enough such that the individual's choice of social activity is higher than the myopic choice determined by (\ref{eq: social_activity_general}) for several periods after $t$ and smaller for all periods after $t+\tau*$.}

\begin{proof} Under rational expectations, the expected utility loss due to the risk of infection, evaluated at time $t$ is:
\begin{equation} 	\mathbb{E}\sum_{\tau = t}^{\infty} \delta^{\tau - t} U^h_\tau = \delta \sigma_t \Big[P^j_{t+1}+ \sum_{\tau=2}^{\infty} \: \delta^{\tau-1} P^j_{t+\tau} \prod_{i=1}^{\tau-1} \Big(1-P^j_{t+i} \Big)     \Big]\alpha \label{EULossRat} \end{equation} 
Comparing Equation ( \ref{EULossRat}) with the corresponding equation for the myopic case, Equation (\ref{EULoss}), it follows that the loss is lower (resp. higher) under myopic expectations if $P^j_{t+\tau}$ is an increasing sequence (resp. decreasing) in $\tau \geq 1$. It follows then that the implied optimal choice of social activity under myopic expectations is higher (resp. lower) if $P^j_{t+\tau}$ is an increasing (resp. decreasing) sequence in $\tau \geq 1$. Finally, a $\tau^*$ large enough and a discount $\delta$ low enough guarantee that the increasing portion of the sequence $P^j_{t+\tau}$ dominates the choice of social action under rational expectations.   \end{proof}

The probability of infection $P^j_{t+1}(\textbf{a}_t, \textbf{i}_t)$, for any agent $j$ in $t$ is characterized in equilibrium as a function of the structure of the network of individuals and their information about the disease. The determination of social actions at equilibrium requires, therefore, closing the model with the network's structure of contacts. In the next section, we specify the substrate over which the disease spreads and the equilibrium condition determining the dynamics of infections.

Finally, denoting $S_t, I_t,R_t$ as the fractions of susceptible, infected, and recovered - respectively - at time $t$, we specify the dynamics of infection in the Susceptible-Infected-Recovered (SIR) model \citep{kermack1927contribution}. \bigskip 

\noindent {\bf Definition: Infection dynamics (SIR).} {\em The Susceptible-Infected-Recovered (SIR) dynamics are characterized by the following dynamical system: 
\begin{subequations}
	\begin{alignat}{1}
		\label{eq: homogeneous_model_s_gen}
		&S_{t+1} = S_t - \beta  P_{t+1} S_t \\
		\label{eq: homogeneous_model_i_gen}
		&I_{t+1} = I_t + \beta  P_{t+1} S_t  - \mu I_t\\
		\label{eq: homogeneous_model_r_gen}
		&R_{t+1} = R_t + \mu I_t 
	\end{alignat}
\end{subequations} where $P_{t+1}$ is the average probability that a susceptible individual is infected at time $t+1$.}

\subsection{Local vs global awareness}
\label{sec:awareness}

In this section, 
we apply the general model analyzed so far to determine how individuals perceive their probability of infection based on social activity. 
We consider various assumptions about the underlying contact network and the information available to individuals within such a network.
For each information structure considered, the simple corollaries of the characterization of social activity in Section \ref{sec:model} will allow us to express each individual's infection probability as a function of their exposure to infectious activity, denoted as $\theta^j_t$.
This variable represents a reduced form representation of the equilibrium social activity of all individuals in the network $(\textbf{a}_t, \textbf{i}_t)$.
This framework allows us to specify the SIR dynamics equations that, at equilibrium, describe the dynamics of the infection. 
 
%

The information structures we study and compare are distinguished in terms of how spread out across the network the information about the infection is. 
We contrast a \emph{global awareness} scenario, where individuals only know the prevalence of the disease in the population – with \emph{local awareness}, where individuals know the prevalence in their neighborhood.
Focusing first on global awareness (Fig. \ref{fig: explicative_fig}d), we distinguish two specific information structures. 
In the first, individuals (represented as nodes in the network) only know the overall prevalence of the disease. 
In the second, individuals know the prevalence within groups of individuals with similar connectivity (i.e., nodes with the same degree). 

Our initial global awareness structure assumes that individuals only know the proportion of susceptible and infected individuals in the entire network. 
Therefore, the perceived probability of being susceptible at time $t$ is simply the total fraction of susceptible individuals $S_t$, and the perceived prevalence is the fraction of infected individuals, $I_t$. 
We further assume that all nodes have the same degree, specifically the average degree of the network ($\langle k \rangle$). 
This is called the homogeneous mean-field (MF) information structure.  
In this context, at each time step $t$, individuals choose their optimal social activity only based on the average prevalence in the population. The behavioral model from Section \ref{sec:model} can then be further characterized as follows. \bigskip

\begin{figure}[tbp]
	\centering
	\includegraphics[width=\textwidth]{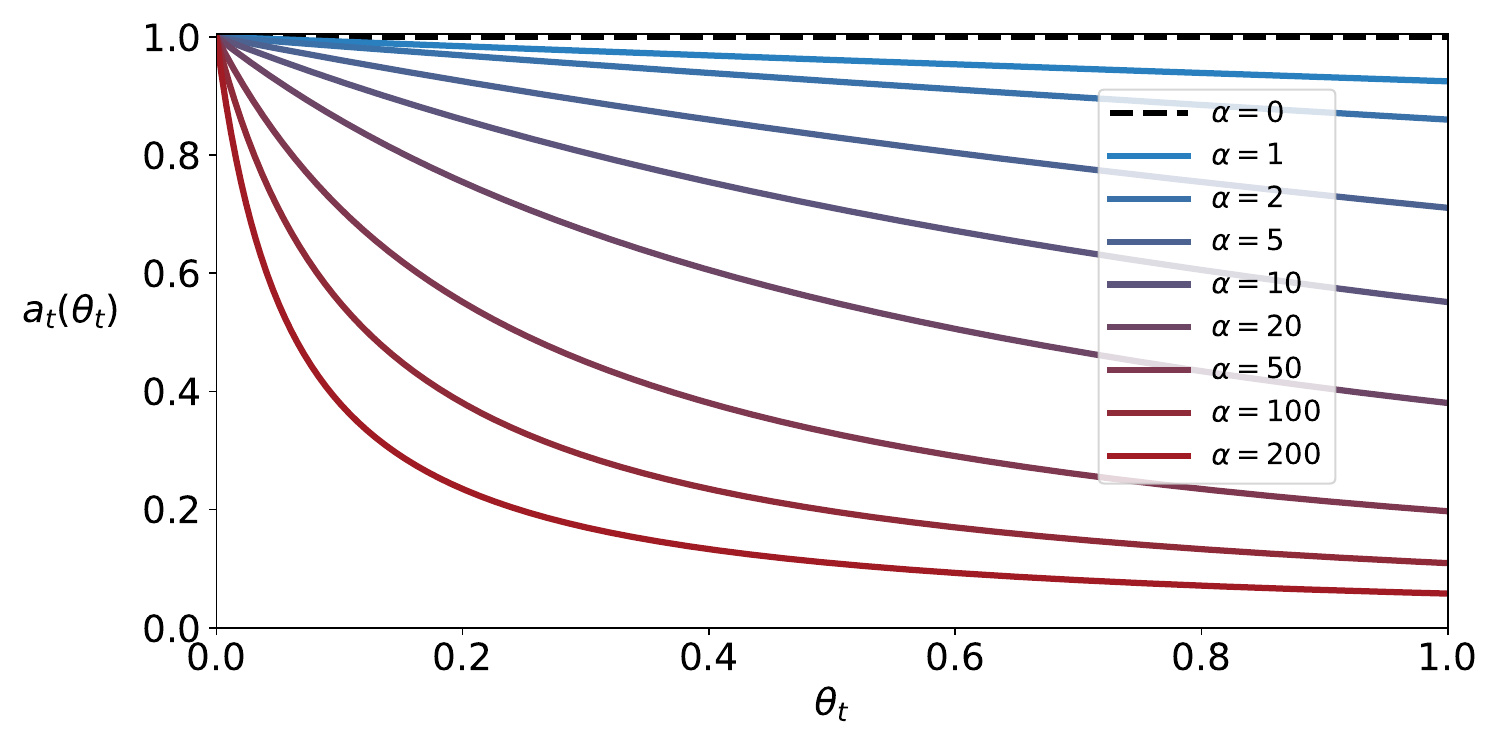}
	\caption{Optimal value of the social activity $a_t$ (Eq. \ref{eq: homogeneous_model_a}) as a function of the aggregate infectious activity $\theta_t$. We fix $\sigma=1$ (corresponding to an individual that is sure of being susceptible), $\delta=0.905$, $\mu = 0.1$, $\beta=0.02$ and $\langle k \rangle = 14.7$. Different colors correspond to different choices of $\alpha$.\label{fig: a_opt}}
\end{figure}

\noindent {\bf Proposition: Social activity and infection dynamics under MF.} 
{\em In the  homogeneous mean-field {(MF)} information structure, each individual (node) adopts the same social activity $a_t$ and perceives a probability of being infected 
$$P^j_{t+1}(\textbf{a}_t, \textbf{i}_t)= P_{t+1}(a_t, \theta_t) =\beta \langle k \rangle a_t \theta_t, $$  where \begin{equation} \theta_t = a_t I_t     \label{eq: homogeneous_model_thetaa} \end{equation}  is the aggregate infectious activity.
Equilibrium social activity, in turn, solves
\begin{equation}
a_t = \frac{1}{1 + \alpha \delta \beta \langle k \rangle S_t \theta_t }. \label{eq: homogeneous_model_a}
\end{equation}    
Consequently, the SIR dynamics of the homogeneous mean-field {(MF)} information structure  are summarized by Equations (\ref{eq: homogeneous_model_thetaa}-\ref{eq: homogeneous_model_a}) and the following:
\begin{subequations}
	\begin{alignat}{1}
		\label{eq: homogeneous_model_s}
		&S_{t+1} = S_t - \beta \langle k \rangle a_t S_t \theta_t\\
		\label{eq: homogeneous_model_i}
		&I_{t+1} = I_t + \beta \langle k \rangle a_t S_t \theta_t - \mu I_t\\
		\label{eq: homogeneous_model_r}
		&R_{t+1} = R_t + \mu I_t 
	\end{alignat}
\end{subequations} }
Fig. \ref{fig: a_opt} shows the optimal social activity as a function of the aggregate infectious activity $\theta$, for different choices of the average cost of infection $\alpha$.
One can see that optimal social activity always decreases as aggregate infectious activity increases, but its functional form depends on $\alpha$: the optimal social activity decreases slowly or more abruptly when the average cost of infection is small or large, respectively.
In particular, when $\alpha$ is small, the optimal social activity decreases linearly with $\alpha$.\footnote{We also tested the effect of the discount factor $\delta$ (not shown), which, in the range of values we consider for $\delta$ (see next Section) is negligible.}

The second global awareness environment we study is referred to as a heterogeneous (degree-based) mean-field structure (HMF) \citep{pastor-satorras2001epidemic}. In this case, the information structure across the network is such that individuals know the degree distribution and prevalence for degree classes, instead of only the aggregate susceptibility and prevalence in the network. 

All individuals (nodes) with the same degree $k$ are equivalent and will thus adopt the same value of social activity $a_{k}$. 
Let $I^{k}_t$ (resp. $S^k_t$) be the fraction of nodes with degree $k$ that are in the infected state $i$ (resp. the susceptible state $s$) at $t$.  
In addition, let $p^k$ denote the fraction of nodes with degree $k$ and  $\langle k \rangle$ the average degree of the network, $\langle k \rangle = \sum_k k p_k$. Within this context, we obtain the following further characterization of social action and of the dynamics of infection at equilibrium.  \bigskip 
 
\noindent {\bf Proposition: Social activity and infection dynamics under HMF. } {\em In the the heterogeneous mean-field {(HMF)} information structure, each individual (node) $j$ with the same degree $k$  adopts the same social activity $a^k_t$ and perceives a probability of being infected 
$$P^j_{t+1}(\textbf{a}_t, \textbf{i}_t)= P^k_{t+1}(a^k_t, \theta_t) = \beta k a^k_{t}  \theta_t, $$ where 
\begin{equation} \theta_t = \sum_{k} a_{k}(k-1)p_k I_t^{k}/\langle k \rangle 	\label{eq: heterogeneous_model_thetaa} \end{equation}  is the aggregate infectious activity. Equilibrium social activity in turn solves \begin{equation}
a^k_t = \frac{1}{1 + \alpha \delta \beta \langle k \rangle S^k_t \theta_t }. \label{eq: heterogeneous_model_a}
\end{equation}   
Consequently, the SIR dynamics of the {heterogeneous mean-field (HMF)} information structure are summarized by Equations (\ref{eq: heterogeneous_model_thetaa}-\ref{eq: heterogeneous_model_a}) and the following:  
\begin{subequations}
	\begin{alignat}{1}
		\label{eq: heterogeneous_model_s}
		&S^k_{t+1} = S^k_{t} - \beta k a^k_{t} S^k_{t} \theta_t\\
		\label{eq: heterogeneous_model_i}
		&I^k_{t+1} = I^k_{t} + \beta k a^k_{t} S^k_{t} \theta_t - \mu I^k_{t}\\
		\label{eq: heterogeneous_model_r}
		&R^k_{t+1} = R^k_{t} + \mu I^k_{t} \\
        &S_t=\sum_{k} p_k S^k_t, \; I_t=\sum_{k} p_k I^k_t, \; R_t=\sum_{k}  p_k R^k_t \label{eq: heterogeneous_model_SIR}
		\end{alignat}
\end{subequations} }
\noindent Prevalence $\theta_t$ in Equation (\ref{eq: heterogeneous_model_thetaa}) is the equivalent of  Equation \eqref{eq: homogeneous_model_thetaa} for the MF structure and is   obtained by weighting the prevalence in each degree class $k$ with its social activity, 
assuming no degree correlations in the contact network. Eq. 		\eqref{eq: heterogeneous_model_a} is the equivalent of Eq. \eqref{eq: homogeneous_model_a} for the MF structure: at each time step $t$, individuals with degree $k$ choose their optimal social activity depending on the aggregate infectious activity $\theta_{t}$. In equilibrium, highly connected individuals (nodes with large degree $k$) will reduce their social activity, {\em ceteris paribus}, with respect to individuals with few social interactions (nodes with small degree $k$).

Lastly, we explore a structure of information of {\it local awareness} (Fig. \ref{fig: explicative_fig}c), in which individuals have detailed knowledge of the local prevalence of their neighborhood.  
Le $\mathcal{N}(j)$ denote the neighborhood of the individual $j$ --- that is the set of nodes $j$ is linked to.
Let the degree, $k^j = |\mathcal{N}(j)|$, indicate the number of neighbors of individual $j$. 
Individuals know the local prevalence $\sum_{n \in \mathcal{N}(j)} \textbf{i}_t^n / k^j$ at time $t$; and they estimate their probability of being susceptible as the fraction of their neighbors currently in the susceptible state $s$, $\sigma^j_t = \sum_{n \in \mathcal{N}(j)} \textbf{s}_t^n / k^j$. \bigskip 

\noindent {\bf Proposition: Social activity under local awareness.} {\em  In the local awareness information structure, each individual (node) adopts a different social activity $a^j_t$ and perceives the probability of being infected 
$$P^j_{t+1}(\textbf{a}_t, \textbf{i}_t)= \beta a^j_k k^j \theta^j_t, $$ where 
\begin{equation} \theta^j_t = \langle a \rangle^j_t \sum_{n \in \mathcal{N}(j)} \textbf{i}_t^n / k^j 	\label{eq: heterogeneous_theta} \end{equation}  is the   infectious activity in the neighborhood of node $j$. 
Equilibrium social activity, in turn, solves \begin{equation}\label{eq: a_opt_quenched}
	a^{j}_t = \frac{1}{1 + \alpha \delta \beta \sigma^j_t k^j \theta^j_t}.
\end{equation} }

\noindent As in the previous mean field information structures, individuals choose their social activity based on their perceived susceptibility, local prevalence, the behavior of their neighbors, infection cost, and discount factor. However, under local awareness, these factors vary across the network. Consequently, the SIR dynamics lack a simple aggregate representation (as seen in the global awareness structures) and must instead be obtained through numerical simulations.
Summarizing, in both global awareness information structures, individuals interact randomly, each potentially coming into contact with all the others. The resulting interaction network is an \textit{annealed} network \citep{pastor-satorras2001epidemic}.
The difference between the MF and the HMF models lies in the fact that the MF model assumes complete homogeneity among individuals, while the HMF model allows for multiple population classes (such as age stratification) and only assumes homogeneity inside each class.

In the case of local awareness, interactions between individuals are completely specified by the network structure, with each node interacting with its neighbors. 
This scenario is typically described by agent-based models with interactions taking place on a \textit{quenched} network \citep{PhysRevLett.105.218701}. 
No assumption of homogeneity is necessary in this case, with all model parameters potentially being different for each individual.
In the local awareness case, we consider two types of underlying network structure: a more homogeneous one (Erdős–Rényi network), that matches the MF case, and a heterogeneous one (scale-free network), closer to the HMF case. 
In this way, we compare different information structures under similar assumptions on the population contact structure.


\subsection{SIS epidemic model}
\label{sec:sis}

To extend our approach to endemic diseases, we incorporate a Susceptible-Infected-Susceptible (SIS) model into our framework. 
In the classic SIS model, infected individuals recover at a constant rate $\mu$ and become susceptible again. 
This dynamic, typical of infections that do not confer long-lasting immunity, leads to a constant prevalence of infectious individuals at equilibrium \citep{keeling2008modeling}.

Under the same assumptions introduced at the beginning of Section \ref{sec:model}, it is possible to derive the closed-form expression objective function at time $t$ as:
\begin{equation}\label{eq: objective_SIS}
	J^j_t = \frac{U(a^j_t)}{1-\delta} - \frac{\alpha \, \delta \, \sigma^j_t(\textbf{s}_t) \, P^j_{t+1} (\textbf{a}_t, \textbf{i}_t)}{1-\delta (1-P^j_{t+1} (\textbf{a}_t, \textbf{i}_t)) - \delta^2 \frac{\mu P^j_{t+1} (\textbf{a}_t, \textbf{i}_t)}{1-\delta(1-\mu)}}.
\end{equation} 

 We note that Eq. \eqref{eq: objective_SIS} is equivalent to Eq. \eqref{eq: objective_simplified}, if we only consider the first order in $\delta$ and a small infection probability $P^j_{t+1} (\textbf{a}_t, \textbf{i}_t) \ll 1$, already assumed to derive Eq. \eqref{eq: objective_simplified}. 
Therefore, we can conclude that in the case of an SIS infection dynamics, the agents will solve the same maximization problem they face with an SIR dynamics, leading to the same expressions for the equilibrium social activity in the local and global awareness scenarios. 
In the following, we will consider the SIR as our main reference model, while also showing results obtained with the SIS model with the same parameters.

\subsection{Model calibration}
\label{sec: behavioral_parameters}

We calibrate the model to fit the context of our analysis: a highly uncertain epidemic environment and a very short-horizon decision problem. We consider a time step equal to one day.\footnote{The code to reproduce the results of the manuscript is available at https://github.com/lorenzoamir/EpiNetworkPaper}
We calibrate the MF model on a disease that has a basic reproduction number $R_0=3$ in the absence of any mitigation measure, implying that a single infection case is expected, on average, to generate three new cases in a population of fully susceptible individuals. 
We set $\mu=1/10$, thus assuming an average infectious period equal to 10 days.
The choices of $R_0$ and $\mu$ imply, for the MF model, a value of the infection rate $\beta=0.3 / \langle k \rangle$, where $\langle k \rangle$ represents the average number of contacts per day. We fix $\langle k \rangle = 14.7$ (see Appendix: Methods for more information on network generation), thus obtaining $\beta = 0.02$.
These epidemiological parameters are compatible with early estimates of $R_0$ and $\mu$ for SARS-CoV-2 \citep{lauer2020incubation,gozzi2021estimating,byrne2020inferred}, and, more in general, with a typical rapidly transmitted respiratory infection, such as influenza \citep{lessler2009outbreak}. 

According to our postulated behavioral rule-of-thumb, individuals believe that current conditions (the sizes of the compartments $s$, $i$ and $r$ and the present value of $\textbf{a}^{-s}_t$) will remain unchanged when planning.
As we noted, we interpret this rule-of-thumb as a short-run high-frequency forecasting strategy. 
Future utility gets exponentially discounted by the term $\delta^{\Delta t}$ in Eq. \eqref{eq: objective}, where $\delta$ is to be interpreted as an ex-ante measure of the agent's perceived precision of the constant epidemic condition assumption, rather than a measure of psychological preference for
the present. Consequently, we set a decay time of 10 days, corresponding to $\delta=\exp\left({-1/10}\right) \simeq 0.905$.\footnote{We test the robustness of the model in the range $\delta \in [0.8, 0.95]$, observing no qualitative changes.} 
With $\delta=0.905$, the utility one week from now weighs about 50\% of today's utility (25\% after two weeks and 5\% after a month). This choice of discount ensures that the planning horizon of individuals is very short.\footnote{When modeling individual choices at relatively low frequencies - especially with rational deterministic expectations about epidemic conditions over longer planning horizons, naturally, discount rates are set to imply higher decay times \citep{fenichel2011adaptive, farboodi2021internal, quaas2021social}.} 
Thus, this assumption has an impact on the plausible values of the discount factor $\delta$, which we choose to be rather small compared to previous modeling approaches \citep{fenichel2011adaptive, farboodi2021internal, quaas2021social}. 





\begin{figure}[tbp]%
\centering
\includegraphics[width=\textwidth]{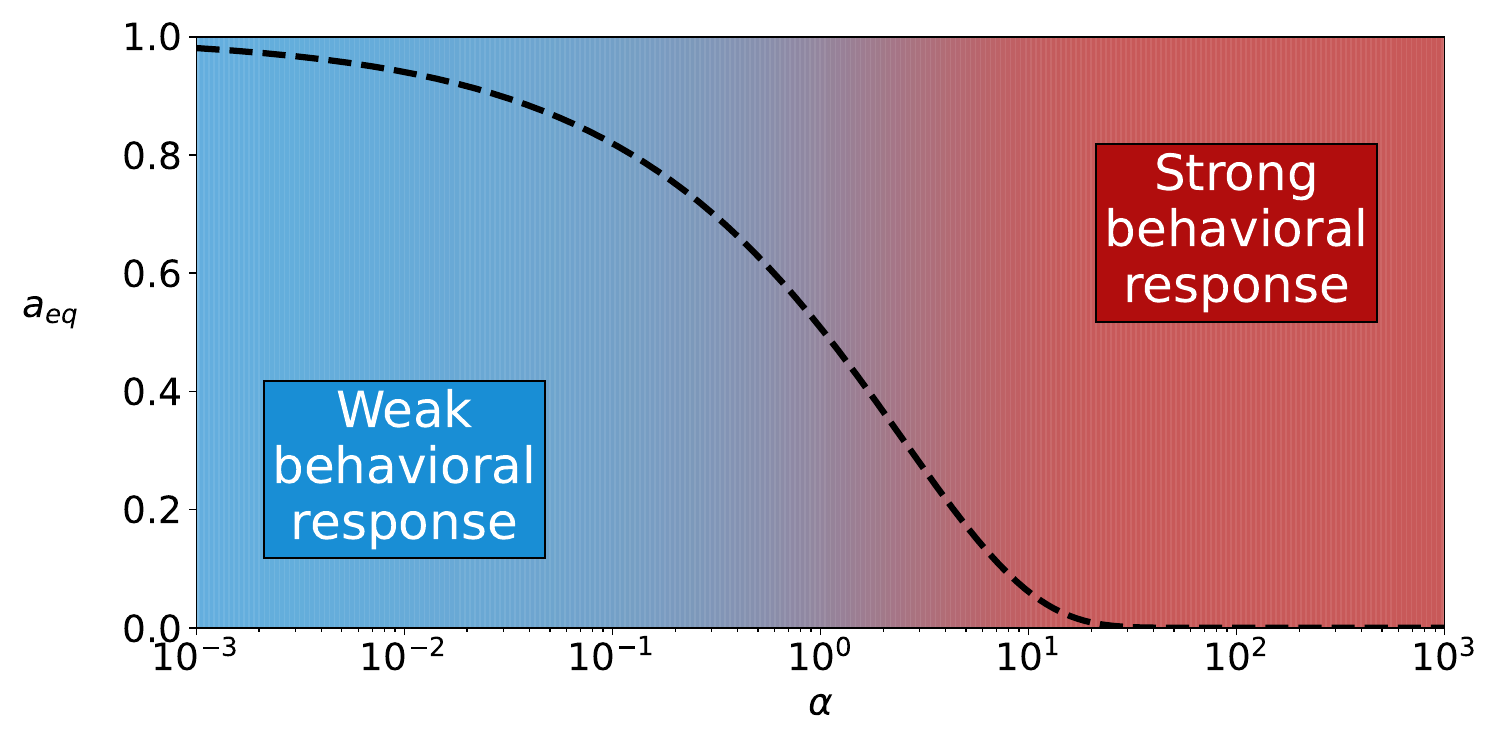}
\caption{Equivalent social activity $a_{eq}$ corresponding to different values of $\alpha$ (dashed line). The dashed line represents the solution of Eq. \eqref{eq:eqiv_soc}. We set $\delta=0.905$ and $\mu=0.1$. The color gradient is based on the values of $a_{eq}$ on the y-axis.}
\label{fig: a_equiv}
\end{figure}

Finally, we address the issue of interpretability of the parameter $\alpha$, which captures an agent's average loss of utility caused by the infection the infection, but lacks a scale to be compared to, that is, we have not yet determined which values of $\alpha$ should be associated with minor illnesses and which with more severe ones.
To this end, we define the equivalent social activity $a_{eq}$ as the value of the social activity for which the utility lost in one day due to the reduction of social activity, $U(a_{eq})$ (given by Eq. \eqref{eq: U_a}), is equal to the utility of an infected individual $U^i$. 
In other words, $a_{eq}$ is such that an individual is indifferent between getting infected and exercising social activity $a_{eq}$. 
This means that, with our choice of the utility function, $a_{eq}$ can be determined numerically as the root of the equation ${log(a_t)-a_t+1+U^i=0}$. 
Using Eq. \eqref{eq: alpha}, we can express the previous equation in terms of $\alpha$, obtaining 
\begin{equation}
\label{eq:eqiv_soc}
 log(a_t)-a_t+1+ \left[1 - \delta(1-\mu) \right]\alpha=0.
\end{equation}

Fig. \ref{fig: a_equiv} shows the equivalent social activity corresponding to different values of $\alpha$ with $\delta=0.905$ and $\mu = 0.1$. 
Intuitively, $a_{eq}$ decreases as the average infection cost increases: individuals are not willing to reduce their social activity if the infection cost is small. 
However, it is interesting to note that the dependency on the average loss of utility $\alpha$ is weak, due to the choice of the logarithmic form of the utility function. 
Individuals are unwilling to significantly decrease their social activity for $\alpha \lessapprox 10^{-2}$, regardless of the probability of infection (prevalence).
They gradually decrease their social activity as a function of $\alpha$, until they are willing to reduce it to almost zero for $\alpha \gtrapprox 10$. 
For these values of $\alpha$, individuals consider becoming infected as serious as having virtually no social activity. 
We stress that this does not necessarily mean that individuals will choose to isolate (i.e., choose zero social activity) to avoid infection, since the probability of getting infected is usually small ($P^j_{t+1} (\textbf{a}_t, \textbf{i}_t) \ll 1$).

With no loss of utility caused by infection, $\alpha=0$, agents do not show a behavioral response. Here, we are interested in the regime for which there is instead a strong behavioral response (highlighted in red in Fig. \ref{fig: a_equiv}). 
Also, we observe that only for high values of $\alpha$ (about $\alpha\simeq100$ in the simulations presented in the next section) behavioral response based on local awareness is strong enough to prevent the epidemic from spreading. 
Therefore, we will focus on the range ${0.1<\alpha<200}$.

\subsection{Numerical simulations}
\label{subsec: numerical_results}
We will now show the results of numerical simulations of the model. 
As described in Section \ref{sec:awareness}, we will consider four settings for the substrate of the epidemic and the information structure. 
The first two represent global awareness: homogeneous mean-field (MF), and heterogeneous mean-field (HMF) on a scale-free network. The last two represent local awareness: Erdős–Rényi (ER) and scale-free (SF) networks. 

\begin{figure}[tbp]%
\centering
\includegraphics[width=\textwidth]{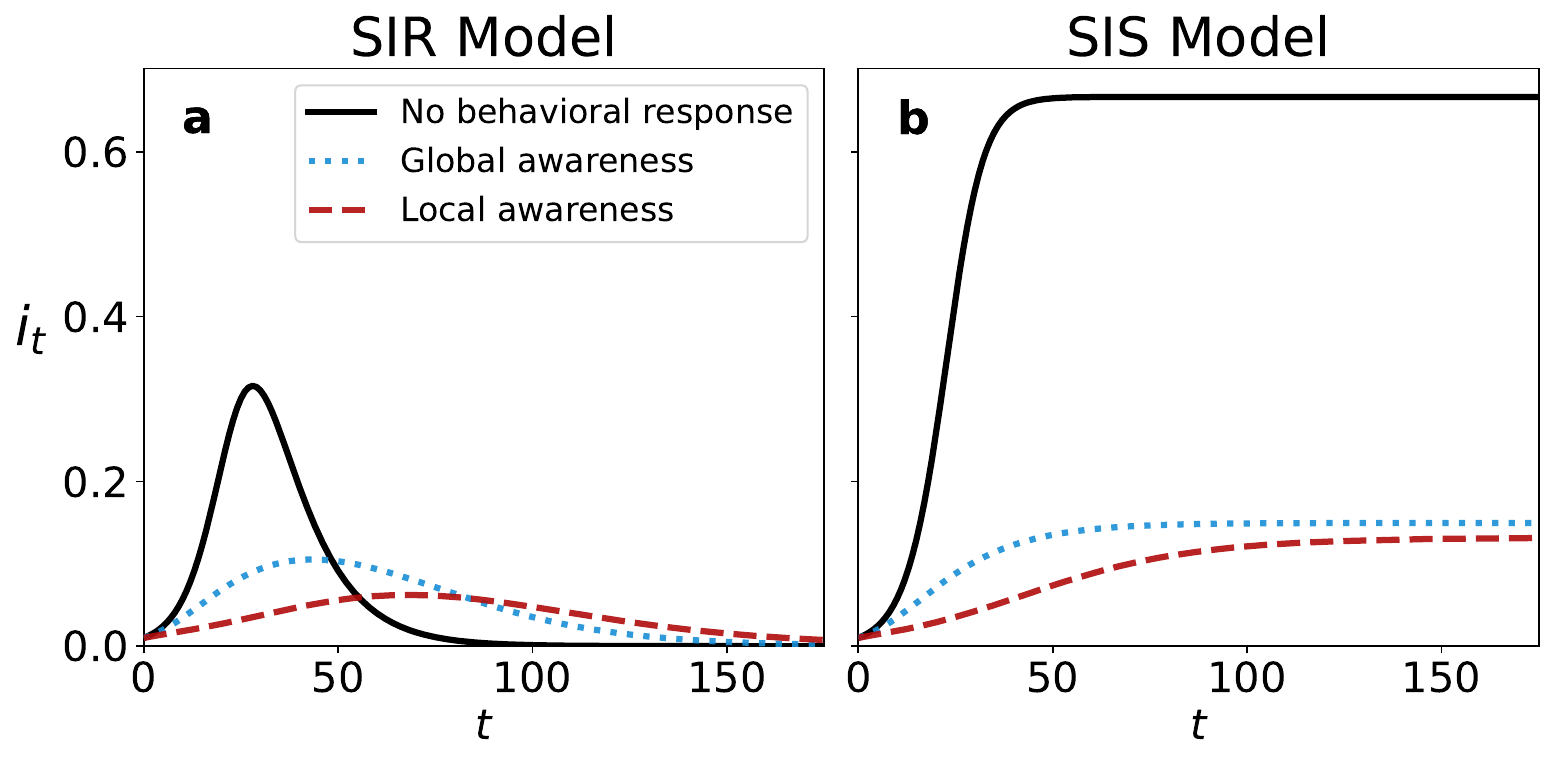}
  \caption{Comparison of local and global awareness scenarios (parametrized by infection cost $\alpha=21$) with no behavioral response ($\alpha=0$). 
  The underlying network is assumed to be MF for global awareness and a Erdős–Rényi network (see Section \ref{sec: network_methods}) for local awareness.
  The top panel shows results for the SIR model, the bottom panel for the SIS model. In both cases, all simulations were performed with $\delta=0.905$, $\mu=0.1$ and $\beta\langle k \rangle = 0.3$} 
\label{fig: epicurves}
\end{figure}

We start by showing the effect of the agents' behavioral response (through forward-looking choices) on the epidemic curve, plotting the prevalence $i_t$ as a function of time in Fig. \ref{fig: epicurves}. 
One can see the epidemic curve of the SIR dynamics is flattened by the behavioral response: with either global or local awareness, the peak of the prevalence is lowered and the infection peak is reached later.
The local awareness scenario seems to be more effective in curbing the disease early than the global awareness scenario. Similar observations hold for the SIS model: the endemic prevalence is much lower with a behavioral response and the equilibrium is reached later in the dynamics. 
Again, local awareness seems slightly more effective than global awareness. 

\begin{figure}[tbp]%
\centering
\includegraphics[width=0.95\textwidth]{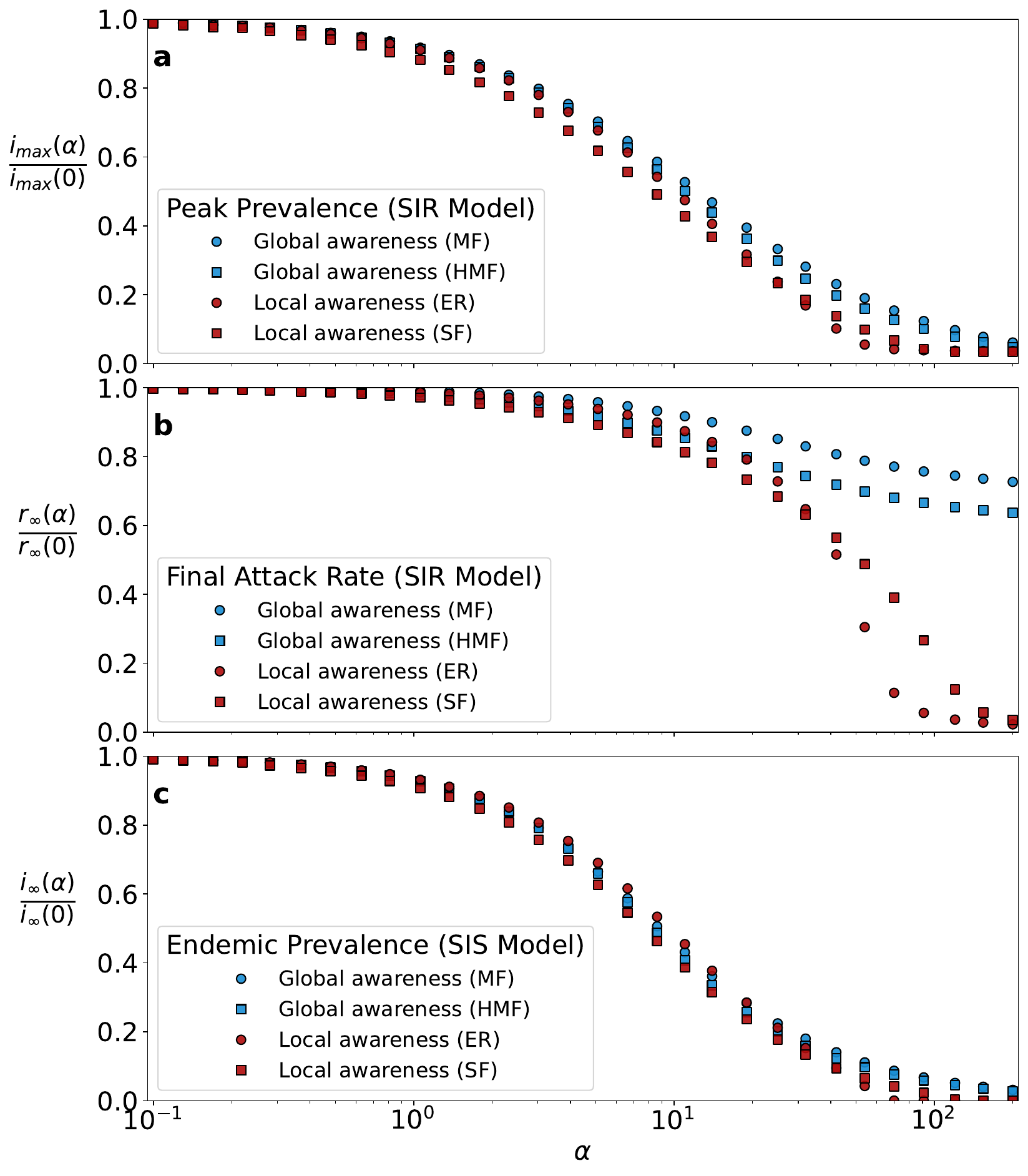}
\caption{Ratio between the peak prevalence $i_{max}$ \textbf{(a)}, the final attack rate $r_{\infty}$ \textbf{(b)} or the SIS equilibrium prevalence $i_{\infty}$ \textbf{(c)} obtained for a given value of $\alpha$ and that corresponding to $\alpha=0$ (no behavioral response).  Error bars are calculated as the standard error of the mean, not shown as smaller than points in the plot.}
\label{fig: subplot}
\end{figure}

We strengthen these observations by measuring the outcome of the epidemic spread via three key quantities: the peak prevalence $i_{max}$ and the {final attack rate} $r_{\infty}$ in the SIR dynamics, and the endemic prevalence $i_{\infty}$ in the SIS dynamics. 
For the SIR model, the first quantity corresponds to the maximum fraction of infected individuals at the same time, the second indicates the fraction of recovered individuals $r_t$ in the limit $t\to\infty$ and represents the total fraction of the population that has been infected by the disease.\footnote{Notice that the peak prevalence can be related to the maximum capacity of the health system at the peak of the epidemic, while the final attack rate can be directly related to the number of deceased individuals.
} 
In the SIS model, $i_{\infty}$ corresponds to the fraction of infected individuals in the endemic phase of the disease, once the equilibrium is reached.

In Fig. \ref{fig: subplot}(a) we plot the peak prevalence $i_{max}$ as a function of $\alpha$, normalized by its value without behavioral response ($\alpha=0$). The plot shows that with a stronger behavioral response (larger $\alpha$) the peak prevalence decreases, indicating that behavioral responses flatten the epidemic curve. 
In Fig. \ref{fig: subplot}(b) we plot how the final attack rate $r_{\infty}$ changes with $\alpha$, also normalized by its value for $\alpha=0$. This plot shows that $r_{\infty}$ decreases as the infection cost increases, for both global and local awareness. 
However, the effects of behavioral change on $r_{\infty}$ are much stronger in the local case than in the global awareness case.
The effect is particularly evident in the regime of strong behavioral response (large $\alpha$ for which $a_{eq}\ll1$). In this case, the epidemic is suppressed when the awareness is local, while if individuals have only global knowledge of the prevalence, the reduction in the final attack rate is relatively small.
Fig. \ref{fig: subplot}(c) shows that the endemic prevalence $i_{\infty}$, normalized by its value for $\alpha=0$, decreases with a stronger behavioral response (larger $\alpha$), in line with the what observed in the SIR model.

The stronger effect of local awareness can be attributed to the more fine-grained behavioral response: people in direct contact with local outbreaks reduce their social activity, hampering early disease propagation. 
Therefore, even when the prevalence is low in the population and is localized in a few individuals, the behavioral response of the neighbors of the infected individuals is sufficient to curb the spread of the disease.
In contrast, within the global setting, the prevalence has to grow enough in the population to trigger a behavioral response from individuals.
The possible presence of clusters (groups of highly connected nodes) in the network acts similarly: when the behavioral response is strong (large $\alpha$ regime), the epidemic cannot escape from the cluster where it started, but it dies after exhausting the reservoir of susceptible individuals within the cluster.
The difference between local and global awareness instead disappears when the behavioral response is weak (small $\alpha$ values). 
In this regime, indeed, a considerable fraction of nodes must be infected before triggering the behavioral response, so that many susceptible individuals will likely be in contact with several infected ones, a situation similar to the global awareness scenario.

\subsection{Policy intervention}

In this Section, we show how our model can capture the interplay between a network-informed policy intervention and individual decision-making. 
We introduce a policy that aims to reduce the social activity of highly connected individuals. We do this indirectly  by adding a term to the utility function that depends linearly on both the individual's social activity and their degree - e.g., a tax or some abtract costly friction on social activity.
This policy is calibrated so that individuals with the highest degree ($k_{max}$) in the network reduce their optimal pre-epidemic social activity by half, while those with the minimum degree ($k_{min}$) maintain their original activity levels.
The general form of the optimal social activity \eqref{eq: social_activity_general} is then

\begin{equation}
    \label{eq: policy_social_activity}
    a^j_t = \frac{
        1
    }
    {
        1 + \frac{k^{j} - k_{min}}{k_{max} - k_{min}} + 
        \alpha \delta \sigma^j_t(\textbf{s}_t)
        \frac{\partial P^j_{t+1}}{\partial a^j_t}
       (\textbf{a}_t, \textbf{i}_t)}, \; \; \forall \; j=1, \ldots, N.
\end{equation}

\begin{figure}[tbp]%
\centering
\includegraphics[width=0.95\textwidth]{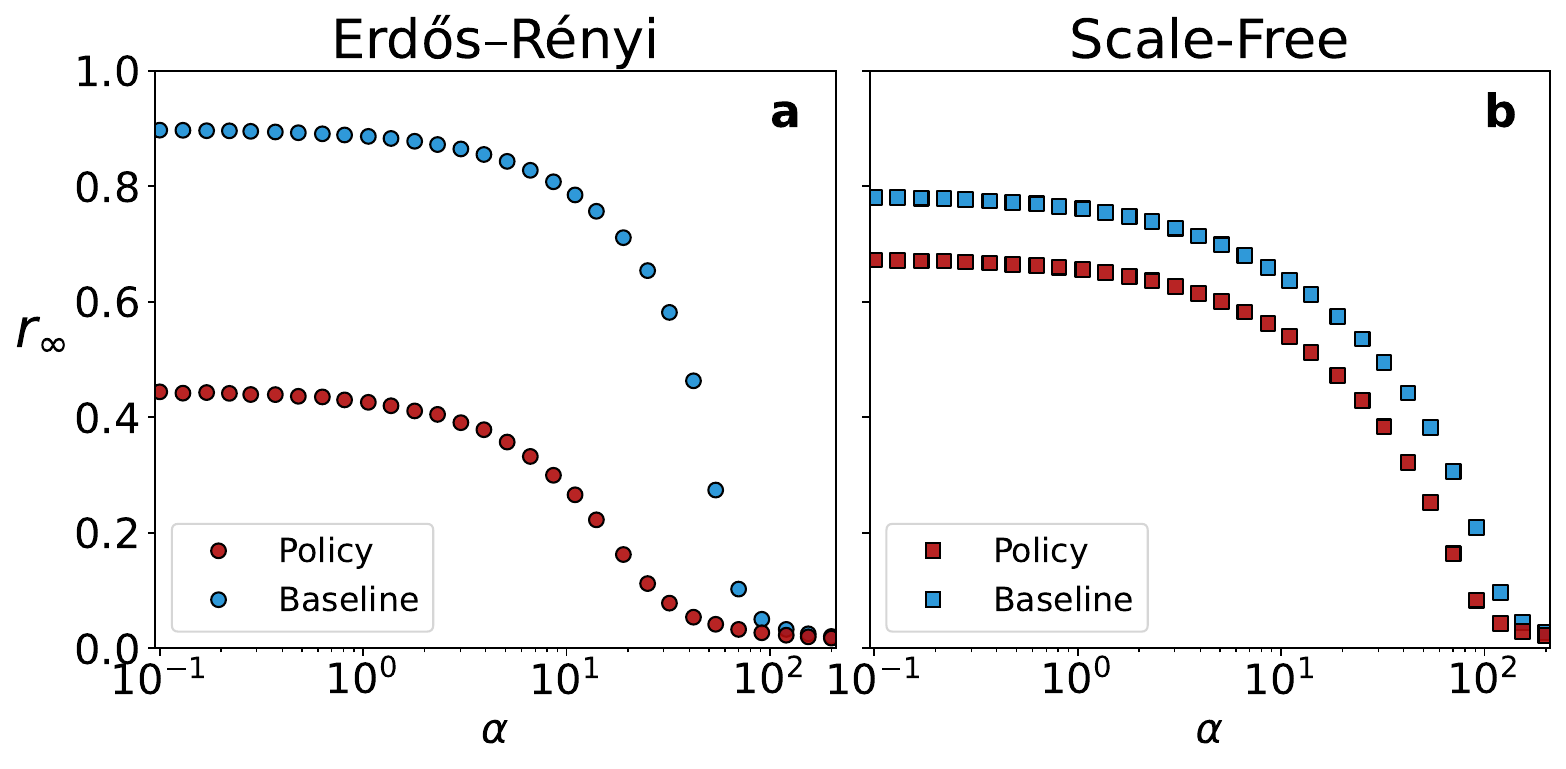}
\caption{Final attack rate $r_{\infty}$ as a function of $\alpha$ in a local awareness scenario with and without policy intervention (Eq. \eqref{eq: policy_social_activity}), on Erdős–Rényi (a) and scale-free (b) networks. 
The baseline against which the policy is evaluated corresponds to the results shown in Fig. \ref{fig: subplot}(b).}
\label{fig: policy}
\end{figure}

Fig. \ref{fig: policy} compares the final attack rate under the policy intervention within the local awareness scenario discussed previously. 
As expected, higher infection costs reduce the final attack rate. 
The policy intervention further decreases this rate, even at low infection costs.
However, degree heterogeneity appears to reduce the policy's impact. For example, at $\alpha=0$ (no perceived infection risk, behavior entirely dictated by policy), the final attack rate is reduced by 50\% in the Erdős–Rényi network, but only by 14\% in the more heterogeneous scale-free network.

Interestingly, the policy intervention synergizes with individual awareness, reducing the level of perceived infection cost needed for significant mitigation.  
To illustrate this point, let us consider the value of $\alpha$ required to decrease the final attack rate, $r_{\infty}(\alpha)$, by 50\% (compared to $r_{\infty}(0)$). 
When the policy is not implemented, these values are relatively similar for the two networks ($\alpha \simeq 42$ Erds-Rényi, $\alpha \simeq 54$ scale-free).
Implementing the policy reduces these values to $\alpha \simeq 14$ and $\alpha \simeq 42$, respectively, making it easier for agents to suppress the spread of the epidemic with their individual decisions.
Note that this effect is more pronounced in the Erdős–Rényi network, highlighting the importance of considering the underlying network substrate when evaluating policy interventions. 

\section{Discussion and Conclusions}
\label{sec:conclusion}

Our study aimed at bridging classical epidemic modeling, in which the transmission rate is modulated by some nonlinear function of the prevalence, and epi-economic models.
These two approaches have been reported to operate largely in isolation and even to show, to some extent, a lack of consensus \citep{darden2022modeling, murray2020epidemiology}.
We adopt the epi-economic framework, in which each individual's behavior is assumed to trade off his/her present gains (from social interactions) and his/her future costs (from the increased probability of infection). 
However, we assume individuals follow a myopic behavioral rule-of-thumb in forecasting future epidemic conditions. 
Crucially, we go beyond the homogeneous mixing hypothesis usually assumed in epi-economic models and test different degrees of heterogeneity in the population considering a class-based approach (global awareness) and an individual-based approach (local awareness).

We show that, under our assumptions, 
it is possible to find an analytical expression for the optimal social activity $a_t$ 
(Eq. \eqref{eq: social_activity_general}), that depends on the infection cost $\alpha$, the discount factor $\delta$, and epidemic conditions that determine the infection probability $P(a^s, \textbf{a}^{-s}, \textbf{i})$ and the probability of being susceptible $\sigma(\textbf{s})$.
While the parameters $\alpha$ and $\delta$ can be calibrated, the probabilities of becoming infected and being susceptible can be estimated in different information scenarios.
Behavioral response models based on \text{prevalence-dependent} transmission rates have been studied 
for decades \citep{capasso1978generalization, wang2012modeling}.
However, in our model, social activity does not depend solely on prevalence but also on the current behavioral choices of the population. Changes in the prevalence are then more relevant when collective behavior favors disease transmission ($a_t\simeq1$) and become less and less relevant as behavior changes limiting the probability of infection ($a_t\ll1$). Moreover, in our model, individuals leverage knowledge of the health status distribution in the population. Consequently, as the disease spreads, they recognize the increasing probability of infection, along with the probability of having acquired immunity following a previous infection.

We operationalize the infection probability considering both homogeneous and heterogeneous behavioral responses, with local or global awareness. 
This approach  enables, as a consequence,  the study of diverse scenarios of heterogeneity, ranging from age stratification to variations in perceived disease severity or planning horizons at single-individual resolution.

We provide a simple interpretation for the infection cost $\alpha$ in terms of an equivalent social activity $a_{eq}$, defined as the acceptable social activity equivalent to the risk of infection. 
This equivalence allows us to distinguish regimes of weak and strong behavioral responses. 
Finally, we quantify the effect of behavior change on the final attack rate of the disease in the strong behavioral response regime, showing that local awareness allows for much stronger outbreak reduction than global awareness.

During the COVID-19 pandemic, a vast amount of empirical measurements of human behavior has become accessible, especially through mobile phone data \citep{oliver2020mobile}.
These have been often used to incorporate human behavior into epidemic models in an effective manner, that is, by retrospectively integrating the observed changes in behavior, for instance, the reduction in movements, into disease dynamics \citep{gozzi2021estimating,di2020impact}.
However, epi-economic models would require rather different, and more granular, empirical measures of human behavior, aimed at quantifying individual future expectations and their heterogeneity in a population. 
Previous studies relied on assumed values for parameters that regulate behavioral responses \citep{ye2021game,farboodi2021internal}. 
However, empirical measurements of these parameters, such as the expected cost of infection, remain scarce.
In this respect it is noteworthy that our model depends only on two key parameters, the infection cost $\alpha$ and the discount factor $\delta$, which makes our model very parsimonious in terms of parameterization needs. 
In general, we remark that data quality is crucial to assess parameters of epidemiological models even in very simple settings, such as the basic reproduction number $R_0$ \citep{starnini_aleta_tizzoni_moreno_2021}.


Our analysis of the effects of including heterogeneous behavioral responses, and in particular the effectiveness of the local awareness scenario, may extend to the case of fully rational, forward-looking agents in a high forecasting precision environment. In fact, despite the different behavioral assumptions, our results are in line with those of \citep{farboodi2021internal}. More precisely, we calibrated our model adopting the parameters of \citep{farboodi2021internal} (which were calibrated on the initial outbreak of COVID-19 in the United States) for our MF model. 
Both models predicted the reproduction number $R(t)$ dropping below 1 in May 2020 and remaining close to $R(t)=1$ afterward. 
The minimum $R(t)$ was $0.947$ in our model and $0.935$ in Ref. \cite{farboodi2021internal}. 
The behavioral response and the share of susceptible individuals were also similar, with both models showing the strongest behavioral response in May 2020 ($a_t=0.6$) and $s_t$ slightly below 0.4 in March 2022. However, our model predicted a stronger reduction in peak prevalence (29\% instead of 25\%) and cumulative fatalities up to July 2020 (83\% instead of 70\%).
We remark that this calibration yields an estimated average cost of infection as $\alpha\simeq 200$, firmly placing it within the realm of strong behavioral response, where the differences between global and local awareness are prominent. 
In conclusion, our study describes a simple yet realistic model of forward-looking behavior that can be integrated into large-scale network epidemic models \citep{chang2021mobility}, contributing an additional layer of realism to models used to inform policymakers. 

Finally, we note that our main result regarding the relative effectiveness of local awareness in flattening an epidemic curve may have relevant policy implications. It suggests that a fine-grained information policy on the part of local media and other institutions may help limit the effects of an epidemic, especially in its early stages. 
\newpage

\bibliography{reference}

\begin{thebibliography}{66}
\providecommand{\natexlab}[1]{#1}
\providecommand{\url}[1]{\texttt{#1}}
\expandafter\ifx\csname urlstyle\endcsname\relax
  \providecommand{\doi}[1]{doi: #1}\else
  \providecommand{\doi}{doi: \begingroup \urlstyle{rm}\Url}\fi

\bibitem[Aadland et~al.(2013)Aadland, Finnoff, and Huang]{aadland2013syphilis}
David Aadland, David~C. Finnoff, and Kevin~X.D. Huang.
\newblock Syphilis cycles.
\newblock \emph{The B.E. Journal of Economic Analysis \& Policy}, 13\penalty0
  (1):\penalty0 297--348, 2013.
\newblock \doi{doi:10.1515/bejeap-2012-0060}.
\newblock URL \url{https://doi.org/10.1515/bejeap-2012-0060}.

\bibitem[Acemoglu et~al.(2020{\natexlab{a}})Acemoglu, Chernozhukov, Werning,
  and Whinston]{acemoglu2020multi}
Daron Acemoglu, Victor Chernozhukov, Iv{\'a}n Werning, and Michael~D Whinston.
\newblock A multi-risk sir model with optimally targeted lockdown.
\newblock National Bureau of Economic Research, 2020{\natexlab{a}}.

\bibitem[Acemoglu et~al.(2020{\natexlab{b}})Acemoglu, Makhdoumi, Malekian, and
  Ozdaglar]{acemoglu2020testing}
Daron Acemoglu, Ali Makhdoumi, Azarakhsh Malekian, and Asuman Ozdaglar.
\newblock Testing, voluntary social distancing and the spread of an infection.
\newblock Technical report, National Bureau of Economic Research,
  2020{\natexlab{b}}.

\bibitem[Aguirregabiria et~al.(2020)Aguirregabiria, Gu, Luo, and Mira]{aguirre}
Victor Aguirregabiria, Jiaying Gu, Yao Luo, and Pedro Mira.
\newblock A dynamic structural model of virus diffusion and network production:
  A first report.
\newblock CEPR Discussion Paper No. DP14750, 2020.

\bibitem[Alfaro et~al.(2020)Alfaro, Faia, Lamersdorf, and
  Saidi]{alfaro2020social}
Laura Alfaro, Ester Faia, Nora Lamersdorf, and Farzad Saidi.
\newblock Social interactions in pandemics: Fear, altruism, and reciprocity.
\newblock Technical report, National Bureau of Economic Research, 2020.

\bibitem[Alvarez et~al.(2021)Alvarez, Argente, and Lippi]{alvarez2021simple}
Fernando Alvarez, David Argente, and Francesco Lippi.
\newblock A simple planning problem for covid-19 lock-down, testing, and
  tracing.
\newblock \emph{American Economic Review: Insights}, 3\penalty0 (3):\penalty0
  367--382, 2021.

\bibitem[Aum et~al.(2020)Aum, Lee, and Shin]{aum2020covid}
Sangmin Aum, Sang Yoon~Tim Lee, and Yongseok Shin.
\newblock Covid-19 doesn't need lockdowns to destroy jobs: The effect of local
  outbreaks in korea.
\newblock Technical report, National Bureau of Economic Research, 2020.

\bibitem[Azzimonti et~al.(2020)Azzimonti, Fogli, Perri, and
  Ponder]{Azzimontietal2020}
Marina Azzimonti, Alessandra Fogli, Fabrizio Perri, and Mark Ponder.
\newblock Pandemic control in econ-epi networks.
\newblock Federal Reserve Bank of Minneapolis Staff report 609, August 2020.

\bibitem[Barrios et~al.(2020)Barrios, Benmelech, Hochberg, Sapienza, and
  Zingales]{barrios}
John~M Barrios, Efraim Benmelech, Yael~V Hochberg, Paola Sapienza, and Luigi
  Zingales.
\newblock Civic capital and social distancing during the covid-19 pandemic.
\newblock Working Paper 27320, National Bureau of Economic Research, June 2020.
\newblock URL \url{http://www.nber.org/papers/w27320}.

\bibitem[Bartik et~al.(2020)Bartik, Bertrand, Cullen, Glaeser, Luca, and
  Stanton]{bartik2020small}
Alexander~W Bartik, Marianne Bertrand, Zo{\"e}~B Cullen, Edward~L Glaeser,
  Michael Luca, and Christopher~T Stanton.
\newblock How are small businesses adjusting to covid-19? early evidence from a
  survey.
\newblock Technical report, National Bureau of Economic Research, 2020.

\bibitem[Batagelj and Brandes(2005)]{batagelj2005efficient}
Vladimir Batagelj and Ulrik Brandes.
\newblock Efficient generation of large random networks.
\newblock \emph{Phys. Rev. E}, 71:\penalty0 036113, Mar 2005.
\newblock \doi{10.1103/PhysRevE.71.036113}.
\newblock URL \url{https://link.aps.org/doi/10.1103/PhysRevE.71.036113}.

\bibitem[Bethune and Korinek(2020)]{bethune2020covid}
Zachary~A Bethune and Anton Korinek.
\newblock Covid-19 infection externalities: Trading off lives vs. livelihoods.
\newblock \emph{National Bureau of Economic Research Working Paper}, 2020.

\bibitem[Bisin and Moro(2022)]{bisin-moro-lockdown-lucas-critique-2022}
Alberto Bisin and Andrea Moro.
\newblock {Spatial-SIR with Network Structure and Behavior: Lockdown Policies
  and the Lucas Critique}.
\newblock \emph{Journal of Economic Behavior \& Organization}, 198:\penalty0
  370--388, June 2022.
\newblock URL \url{http://andreamoro.net/assets/papers/LockdownLC.pdf}.

\bibitem[Brotherhood et~al.(2020)Brotherhood, Kircher, Santos, and
  Tertilt]{brotherhood2020economic}
Luiz Brotherhood, Philipp Kircher, Cezar Santos, and Mich{\`e}le Tertilt.
\newblock An economic model of the covid-19 epidemic: The importance of testing
  and age-specific policies.
\newblock \emph{CEPR Discussion Paper No. DP14695}, 2020.

\bibitem[Byrne et~al.(2020)Byrne, McEvoy, Collins, Hunt, Casey, Barber, Butler,
  Griffin, Lane, McAloon, et~al.]{byrne2020inferred}
Andrew~William Byrne, David McEvoy, Aine~B Collins, Kevin Hunt, Miriam Casey,
  Ann Barber, Francis Butler, John Griffin, Elizabeth~A Lane, Conor McAloon,
  et~al.
\newblock Inferred duration of infectious period of sars-cov-2: rapid scoping
  review and analysis of available evidence for asymptomatic and symptomatic
  covid-19 cases.
\newblock \emph{BMJ open}, 10\penalty0 (8):\penalty0 e039856, 2020.

\bibitem[Capasso and Serio(1978)]{capasso1978generalization}
Vincenzo Capasso and Gabriella Serio.
\newblock A generalization of the kermack-mckendrick deterministic epidemic
  model.
\newblock \emph{Mathematical Biosciences}, 42\penalty0 (1):\penalty0 43--61,
  1978.
\newblock ISSN 0025-5564.
\newblock \doi{https://doi.org/10.1016/0025-5564(78)90006-8}.
\newblock URL
  \url{https://www.sciencedirect.com/science/article/pii/0025556478900068}.

\bibitem[Castellano and Pastor-Satorras(2010)]{PhysRevLett.105.218701}
Claudio Castellano and Romualdo Pastor-Satorras.
\newblock Thresholds for epidemic spreading in networks.
\newblock \emph{Phys. Rev. Lett.}, 105:\penalty0 218701, Nov 2010.
\newblock \doi{10.1103/PhysRevLett.105.218701}.
\newblock URL \url{https://link.aps.org/doi/10.1103/PhysRevLett.105.218701}.

\bibitem[Chang et~al.(2021)Chang, Pierson, Koh, Gerardin, Redbird, Grusky, and
  Leskovec]{chang2021mobility}
Serina Chang, Emma Pierson, Pang~Wei Koh, Jaline Gerardin, Beth Redbird, David
  Grusky, and Jure Leskovec.
\newblock Mobility network models of covid-19 explain inequities and inform
  reopening.
\newblock \emph{Nature}, 589\penalty0 (7840):\penalty0 82--87, 2021.

\bibitem[Coibion et~al.(2020)Coibion, Gorodnichenko, and
  Weber]{coibion2020cost}
Olivier Coibion, Yuriy Gorodnichenko, and Michael Weber.
\newblock The cost of the covid-19 crisis: Lockdowns, macroeconomic
  expectations, and consumer spending.
\newblock Technical report, National Bureau of Economic Research, 2020.

\bibitem[Crosby(2003)]{crosby2003america}
Alfred~W. Crosby.
\newblock \emph{{America's Forgotten Pandemic: The Influenza of 1918}}.
\newblock Cambridge University Press, 2 edition, 2003.
\newblock \doi{10.1017/CBO9780511586576}.

\bibitem[Darden et~al.(2022)Darden, Dowdy, Gardner, Hamilton, Kopecky, Marx,
  Papageorge, Polsky, Powers, Stuart, and Zahn]{darden2022modeling}
Michael~E. Darden, David Dowdy, Lauren Gardner, Barton~H. Hamilton, Karen
  Kopecky, Melissa Marx, Nicholas~W. Papageorge, Daniel Polsky, Kimberly~A.
  Powers, Elizabeth~A. Stuart, and Matthew~V. Zahn.
\newblock {Modeling to inform economy-wide pandemic policy: Bringing
  epidemiologists and economists together}.
\newblock \emph{Health Economics}, 31\penalty0 (7):\penalty0 1291--1295, 2022.
\newblock \doi{https://doi.org/10.1002/hec.4527}.
\newblock URL \url{https://onlinelibrary.wiley.com/doi/abs/10.1002/hec.4527}.

\bibitem[Di~Domenico et~al.(2020)Di~Domenico, Pullano, Sabbatini, Bo{\"e}lle,
  and Colizza]{di2020impact}
Laura Di~Domenico, Giulia Pullano, Chiara~E Sabbatini, Pierre-Yves Bo{\"e}lle,
  and Vittoria Colizza.
\newblock Impact of lockdown on covid-19 epidemic in {\^i}le-de-france and
  possible exit strategies.
\newblock \emph{BMC medicine}, 18\penalty0 (1):\penalty0 1--13, 2020.

\bibitem[d’Andrea et~al.(2022)d’Andrea, Gallotti, Castaldo, and
  De~Domenico]{d2022individual}
Valeria d’Andrea, Riccardo Gallotti, Nicola Castaldo, and Manlio De~Domenico.
\newblock Individual risk perception and empirical social structures shape the
  dynamics of infectious disease outbreaks.
\newblock \emph{PLoS Computational Biology}, 18\penalty0 (2):\penalty0
  e1009760, 2022.

\bibitem[Epstein et~al.(2008)Epstein, Parker, Cummings, and
  Hammond]{epstein2008coupled}
Joshua~M Epstein, Jon Parker, Derek Cummings, and Ross~A Hammond.
\newblock Coupled contagion dynamics of fear and disease: mathematical and
  computational explorations.
\newblock \emph{PloS one}, 3\penalty0 (12):\penalty0 e3955, 2008.

\bibitem[Farboodi et~al.(2021)Farboodi, Jarosch, and
  Shimer]{farboodi2021internal}
Maryam Farboodi, Gregor Jarosch, and Robert Shimer.
\newblock Internal and external effects of social distancing in a pandemic.
\newblock \emph{Journal of Economic Theory}, 196:\penalty0 105293, 2021.
\newblock ISSN 0022-0531.
\newblock \doi{https://doi.org/10.1016/j.jet.2021.105293}.
\newblock URL
  \url{https://www.sciencedirect.com/science/article/pii/S0022053121001101}.

\bibitem[Fenichel et~al.(2011)Fenichel, Castillo-Chavez, Ceddia, Chowell,
  Gonzalez~Parra, Hickling, Holloway, Horan, Morin, Perrings, Springborn,
  Velazquez, and Villalobos]{fenichel2011adaptive}
Eli~P. Fenichel, Carlos Castillo-Chavez, M.~G. Ceddia, Gerardo Chowell,
  Paula~A. Gonzalez~Parra, Graham~J. Hickling, Garth Holloway, Richard Horan,
  Benjamin Morin, Charles Perrings, Michael Springborn, Leticia Velazquez, and
  Cristina Villalobos.
\newblock Adaptive human behavior in epidemiological models.
\newblock \emph{Proceedings of the National Academy of Sciences}, 2011.
\newblock ISSN 0027-8424.
\newblock \doi{10.1073/pnas.1011250108}.
\newblock URL \url{https://www.pnas.org/content/early/2011/03/21/1011250108}.

\bibitem[Ferguson(2007)]{ferguson2007capturing}
Neil~M. Ferguson.
\newblock Capturing human behaviour.
\newblock \emph{Nature}, 446:\penalty0 733--733, 2007.

\bibitem[Funk et~al.(2009)Funk, Gilad, Watkins, and Jansen]{funk2009spread}
Sebastian Funk, Erez Gilad, Chris Watkins, and Vincent~AA Jansen.
\newblock The spread of awareness and its impact on epidemic outbreaks.
\newblock \emph{Proceedings of the National Academy of Sciences}, 106\penalty0
  (16):\penalty0 6872--6877, 2009.

\bibitem[Funk et~al.(2010)Funk, Salath{\'e}, and Jansen]{funk2010modelling}
Sebastian Funk, Marcel Salath{\'e}, and Vincent~AA Jansen.
\newblock Modelling the influence of human behaviour on the spread of
  infectious diseases: a review.
\newblock \emph{Journal of the Royal Society Interface}, 7\penalty0
  (50):\penalty0 1247--1256, 2010.

\bibitem[Funk et~al.(2015)Funk, Bansal, Bauch, Eames, Edmunds, Galvani, and
  Klepac]{funk2015nine}
Sebastian Funk, Shweta Bansal, Chris~T Bauch, Ken~TD Eames, W~John Edmunds,
  Alison~P Galvani, and Petra Klepac.
\newblock Nine challenges in incorporating the dynamics of behaviour in
  infectious diseases models.
\newblock \emph{Epidemics}, 10:\penalty0 21--25, 2015.

\bibitem[Geoffard and Philipson(1996)]{geoffard1996rational}
Pierre-Yves Geoffard and Tomas Philipson.
\newblock Rational epidemics and their public control.
\newblock \emph{International economic review}, pages 603--624, 1996.

\bibitem[Goenka and Liu(2012)]{goenka2012infectious}
Aditya Goenka and Lin Liu.
\newblock Infectious diseases and endogenous fluctuations.
\newblock \emph{Economic Theory}, 50\penalty0 (1):\penalty0 125--149, 2012.

\bibitem[Gonzalez-Eiras and Niepelt(2023)]{gonzalez2023optimal}
Martin Gonzalez-Eiras and Dirk Niepelt.
\newblock Optimal epidemic control.
\newblock 2023.

\bibitem[Goolsbee and Syverson(2020)]{goolsbee2020fear}
Austan Goolsbee and Chad Syverson.
\newblock Fear, lockdown, and diversion: comparing drivers of pandemic economic
  decline 2020.
\newblock Technical report, National Bureau of Economic Research, 2020.

\bibitem[Gozzi et~al.(2021)Gozzi, Tizzoni, Chinazzi, Ferres, Vespignani, and
  Perra]{gozzi2021estimating}
Nicol{\`o} Gozzi, Michele Tizzoni, Matteo Chinazzi, Leo Ferres, Alessandro
  Vespignani, and Nicola Perra.
\newblock Estimating the effect of social inequalities on the mitigation of
  covid-19 across communities in santiago de chile.
\newblock \emph{Nature communications}, 12\penalty0 (1):\penalty0 1--9, 2021.

\bibitem[Granell et~al.(2013)Granell, G\'omez, and
  Arenas]{granell2013dynamical}
Clara Granell, Sergio G\'omez, and Alex Arenas.
\newblock Dynamical interplay between awareness and epidemic spreading in
  multiplex networks.
\newblock \emph{Phys. Rev. Lett.}, 111:\penalty0 128701, Sep 2013.
\newblock \doi{10.1103/PhysRevLett.111.128701}.
\newblock URL \url{https://link.aps.org/doi/10.1103/PhysRevLett.111.128701}.

\bibitem[Gupta et~al.(2020)Gupta, Montenovo, Nguyen, Rojas, Schmutte, Simon,
  Weinberg, and Wing]{gupta2020effects}
Sumedha Gupta, Laura Montenovo, Thuy~D Nguyen, Felipe~Lozano Rojas, Ian~M
  Schmutte, Kosali~I Simon, Bruce~A Weinberg, and Coady Wing.
\newblock Effects of social distancing policy on labor market outcomes.
\newblock Technical report, National Bureau of Economic Research, 2020.

\bibitem[Hagberg et~al.(2008)Hagberg, Schult, and Swart]{aric2008exploring}
Aric~A. Hagberg, Daniel~A. Schult, and Pieter~J. Swart.
\newblock Exploring network structure, dynamics, and function using networkx.
\newblock In Ga\"el Varoquaux, Travis Vaught, and Jarrod Millman, editors,
  \emph{Proceedings of the 7th Python in Science Conference}, pages 11 -- 15,
  Pasadena, CA USA, 2008.

\bibitem[Jones and Salathé(2009)]{jones2009early}
James Jones and Marcel Salathé.
\newblock Early assessment of anxiety and behavioral response to novel
  swine-origin influenza a(h1n1).
\newblock \emph{PloS one}, 4:\penalty0 e8032, 12 2009.
\newblock \doi{10.1371/journal.pone.0008032}.

\bibitem[Kahn et~al.(2020)Kahn, Lange, and Wiczer]{NBERw27061}
Lisa~B Kahn, Fabian Lange, and David~G Wiczer.
\newblock Labor demand in the time of covid-19: Evidence from vacancy postings
  and ui claims.
\newblock Working Paper 27061, National Bureau of Economic Research, April
  2020.
\newblock URL \url{http://www.nber.org/papers/w27061}.

\bibitem[Keeling and Rohani(2008)]{keeling2008modeling}
Matt~J Keeling and Pejman Rohani.
\newblock \emph{Modeling infectious diseases in humans and animals}.
\newblock Princeton university press, 2008.

\bibitem[Keppo et~al.(2020)Keppo, Kudlyak, Quercioli, Smith, and
  Wilson]{keppo2020behavioral}
Juusi Keppo, Marianna Kudlyak, Elena Quercioli, Lones Smith, and Andrea Wilson.
\newblock The behavioral sir model, with applications to the swine flu and
  covid-19 pandemics.
\newblock In \emph{Virtual Macro Seminar}, 2020.

\bibitem[Kermack and McKendrick(1927)]{kermack1927contribution}
William~Ogilvy Kermack and Anderson~G McKendrick.
\newblock A contribution to the mathematical theory of epidemics.
\newblock \emph{Proceedings of the royal society of london. Series A,
  Containing papers of a mathematical and physical character}, 115\penalty0
  (772):\penalty0 700--721, 1927.

\bibitem[Kim et~al.(2017)Kim, Cheon, Choi, Joh, and Lee]{kim2017exposure}
Chansung Kim, Seung~Hoon Cheon, Keechoo Choi, Chang-Hyeon Joh, and Hyuk-Jin
  Lee.
\newblock Exposure to fear: Changes in travel behavior during mers outbreak in
  seoul.
\newblock \emph{KSCE Journal of Civil Engineering}, 21\penalty0 (7):\penalty0
  2888--2895, 2017.

\bibitem[Lauer et~al.(2020)Lauer, Grantz, Bi, Jones, Zheng, Meredith, Azman,
  Reich, and Lessler]{lauer2020incubation}
Stephen~A Lauer, Kyra~H Grantz, Qifang Bi, Forrest~K Jones, Qulu Zheng,
  Hannah~R Meredith, Andrew~S Azman, Nicholas~G Reich, and Justin Lessler.
\newblock The incubation period of coronavirus disease 2019 (covid-19) from
  publicly reported confirmed cases: estimation and application.
\newblock \emph{Annals of internal medicine}, 172\penalty0 (9):\penalty0
  577--582, 2020.

\bibitem[Lessler et~al.(2009)Lessler, Reich, Cummings, of~Health, and
  Team]{lessler2009outbreak}
Justin Lessler, Nicholas~G Reich, Derek~AT Cummings, New York City~Department
  of~Health, and Mental Hygiene Swine Influenza~Investigation Team.
\newblock Outbreak of 2009 pandemic influenza a (h1n1) at a new york city
  school.
\newblock \emph{New England Journal of Medicine}, 361\penalty0 (27):\penalty0
  2628--2636, 2009.

\bibitem[Manfredi and D'Onofrio(2013)]{manfredi2013modeling}
Piero Manfredi and Alberto D'Onofrio.
\newblock \emph{Modeling the interplay between human behavior and the spread of
  infectious diseases}.
\newblock Springer Science \& Business Media, 2013.

\bibitem[McAdams(2021)]{mcadams2021blossoming}
David McAdams.
\newblock The blossoming of economic epidemiology.
\newblock \emph{Annual Review of Economics}, 13:\penalty0 539--570, 2021.

\bibitem[Meyers et~al.(2005)Meyers, Pourbohloul, Newman, Skowronski, and
  Brunham]{meyers2005network}
Lauren~Ancel Meyers, Babak Pourbohloul, Mark~EJ Newman, Danuta~M Skowronski,
  and Robert~C Brunham.
\newblock Network theory and sars: predicting outbreak diversity.
\newblock \emph{Journal of theoretical biology}, 232\penalty0 (1):\penalty0
  71--81, 2005.

\bibitem[Murray(2020)]{murray2020epidemiology}
Eleanor~J. Murray.
\newblock {Epidemiology's Time of Need: COVID-19 Calls for Epidemic-Related
  Economics}.
\newblock \emph{Journal of Economic Perspectives}, 34\penalty0 (4):\penalty0
  105--20, November 2020.
\newblock \doi{10.1257/jep.34.4.105}.
\newblock URL \url{https://www.aeaweb.org/articles?id=10.1257/jep.34.4.105}.

\bibitem[Nardin et~al.(2016)Nardin, Miller, Ridenhour, Krone, Joyce, and
  Baumgaertner]{nardin2016planning}
Luis~G Nardin, Craig~R Miller, Benjamin~J Ridenhour, Stephen~M Krone, Paul
  Joyce, and Bert~O Baumgaertner.
\newblock Planning horizon affects prophylactic decision-making and epidemic
  dynamics.
\newblock \emph{PeerJ}, 4:\penalty0 e2678, 2016.

\bibitem[Newman(2003)]{newman2003structure}
M.~E.~J. Newman.
\newblock The structure and function of complex networks.
\newblock \emph{SIAM Review}, 45\penalty0 (2):\penalty0 167--256, 2003.
\newblock \doi{10.1137/S003614450342480}.
\newblock URL \url{https://doi.org/10.1137/S003614450342480}.

\bibitem[Oliver et~al.(2020)Oliver, Lepri, Sterly, Lambiotte, Deletaille,
  De~Nadai, Letouz{\'e}, Salah, Benjamins, Cattuto, et~al.]{oliver2020mobile}
Nuria Oliver, Bruno Lepri, Harald Sterly, Renaud Lambiotte, S{\'e}bastien
  Deletaille, Marco De~Nadai, Emmanuel Letouz{\'e}, Albert~Ali Salah, Richard
  Benjamins, Ciro Cattuto, et~al.
\newblock Mobile phone data for informing public health actions across the
  covid-19 pandemic life cycle, 2020.

\bibitem[Pastor-Satorras and Vespignani(2001)]{pastor-satorras2001epidemic}
Romualdo Pastor-Satorras and Alessandro Vespignani.
\newblock Epidemic spreading in scale-free networks.
\newblock \emph{Phys. Rev. Lett.}, 86:\penalty0 3200--3203, Apr 2001.
\newblock \doi{10.1103/PhysRevLett.86.3200}.
\newblock URL \url{https://link.aps.org/doi/10.1103/PhysRevLett.86.3200}.

\bibitem[Perra et~al.(2011)Perra, Balcan, Gon{\c{c}}alves, and
  Vespignani]{perra2011towards}
Nicola Perra, Duygu Balcan, Bruno Gon{\c{c}}alves, and Alessandro Vespignani.
\newblock Towards a characterization of behavior-disease models.
\newblock \emph{PloS one}, 6\penalty0 (8):\penalty0 e23084, 2011.

\bibitem[Phelan and Toda(2022)]{phelan2022optimal}
Thomas Phelan and Alexis~Akira Toda.
\newblock Optimal epidemic control in equilibrium with imperfect testing and
  enforcement.
\newblock \emph{Journal of Economic Theory}, 206:\penalty0 105570, 2022.

\bibitem[Poletti et~al.(2012)Poletti, Ajelli, and Merler]{poletti2012risk}
Piero Poletti, Marco Ajelli, and Stefano Merler.
\newblock Risk perception and effectiveness of uncoordinated behavioral
  responses in an emerging epidemic.
\newblock \emph{Mathematical Biosciences}, 238\penalty0 (2):\penalty0 80--89,
  2012.

\bibitem[Quaas et~al.(2021)Quaas, Meya, Schenk, Bos, Drupp, and
  Requate]{quaas2021social}
Martin~F. Quaas, Jasper~N. Meya, Hanna Schenk, Björn Bos, Moritz~A. Drupp, and
  Till Requate.
\newblock The social cost of contacts: Theory and evidence for the first wave
  of the covid-19 pandemic in germany.
\newblock \emph{PLOS ONE}, 16\penalty0 (3):\penalty0 1--29, 03 2021.
\newblock \doi{10.1371/journal.pone.0248288}.
\newblock URL \url{https://doi.org/10.1371/journal.pone.0248288}.

\bibitem[Rojas et~al.(2020)Rojas, Jiang, Montenovo, Simon, Weinberg, and
  Wing]{rojas2020cure}
Felipe~Lozano Rojas, Xuan Jiang, Laura Montenovo, Kosali~I Simon, Bruce~A
  Weinberg, and Coady Wing.
\newblock Is the cure worse than the problem itself? immediate labor market
  effects of covid-19 case rates and school closures in the us.
\newblock Technical report, National Bureau of Economic Research, 2020.

\bibitem[Rowthorn and Toxvaerd(2012)]{rowthorn2012optimal}
Bob~RE Rowthorn and Flavio Toxvaerd.
\newblock The optimal control of infectious diseases via prevention and
  treatment.
\newblock CEPR Discussion Paper No. DP8925, 2012.

\bibitem[Starnini et~al.(2021)Starnini, Aleta, Tizzoni, and
  Moreno]{starnini_aleta_tizzoni_moreno_2021}
Michele Starnini, Alberto Aleta, Michele Tizzoni, and Yamir Moreno.
\newblock Impact of data accuracy on the evaluation of covid-19 mitigation
  policies.
\newblock \emph{Data \& Policy}, 3:\penalty0 e28, 2021.
\newblock \doi{10.1017/dap.2021.25}.

\bibitem[Tirupathi et~al.(2020)Tirupathi, Bharathidasan, Palabindala, Salim,
  and Al-Tawfiq]{Tirupathi2020comprehensive}
Raghavendra Tirupathi, Kavya Bharathidasan, Venkataraman Palabindala,
  Sohail~Abdul Salim, and Jaffar~A Al-Tawfiq.
\newblock Comprehensive review of mask utility and challenges during the
  covid-19 pandemic.
\newblock \emph{Infez Med}, 28\penalty0 (suppl 1):\penalty0 57--63, 2020.

\bibitem[Toxvaerd(2020)]{toxvaerd2020equilibrium}
FMO Toxvaerd.
\newblock Equilibrium social distancing.
\newblock Faculty of Economics, University of Cambridge, 2020.

\bibitem[Verelst et~al.(2016)Verelst, Willem, and
  Beutels]{verelst2016behavioural}
Frederik Verelst, Lander Willem, and Philippe Beutels.
\newblock Behavioural change models for infectious disease transmission: a
  systematic review (2010--2015).
\newblock \emph{Journal of The Royal Society Interface}, 13\penalty0
  (125):\penalty0 20160820, 2016.

\bibitem[Wang(2012)]{wang2012modeling}
W.~Wang.
\newblock Modeling adaptive behavior in influenza transmission.
\newblock \emph{Mathematical Modelling of Natural Phenomena}, 7\penalty0
  (3):\penalty0 253–262, 2012.
\newblock \doi{10.1051/mmnp/20127315}.

\bibitem[Ye et~al.(2021)Ye, Zino, Rizzo, and Cao]{ye2021game}
Mengbin Ye, Lorenzo Zino, Alessandro Rizzo, and Ming Cao.
\newblock Game-theoretic modeling of collective decision making during
  epidemics.
\newblock \emph{Physical Review E}, 104\penalty0 (2):\penalty0 024314, 2021.

\end{thebibliography}

\newpage 
\appendix

\section*{Appendix: Methods} \label{sec: network_methods}

Here, we provide details about how we carried out numerical simulations.
Simulations on quenched networks (both the Erdős–Rényi and the scale-free case) are performed by looping over infected nodes and their neighbors and choosing whether transitions happen or not based on generating a random variable.
At each time step, the contagion occurs in each connected pair of one $s$ node and $i$ node with probability $\beta$ and each $i$ individual has a probability $\mu$ of recovery. 
Mean-field (MF) simulations  are based on equations   \ref{eq: homogeneous_model_thetaa}- \ref{eq: homogeneous_model_a} and   \ref{eq: homogeneous_model_s} - \ref{eq: homogeneous_model_r}; Mean field    (HMF) are based on \ref{eq: heterogeneous_model_thetaa} - \ref{eq: heterogeneous_model_a} and \ref{eq: heterogeneous_model_s} - \ref{eq: heterogeneous_model_SIR}.
All simulations presented in this manuscript are performed with $\beta\langle k \rangle=0.3$, $\mu=0.1$, and initial conditions ${I_0=0.01}$ and $a_0=1$ for all nodes. 

For simulations on quenched networks (local awareness), results correspond to ensemble averages over $10^3$ simulations (and thus $10^3$ quenched networks). 
All networks have $N=10^4$ nodes.
To exclude finite size effects, simulations with $N = 10^3$ and $N = 10^5$ nodes have been performed too. 
Scale-free networks are generated using the configuration model \citep{newman2003structure} implemented in the NetworkX Python module \citep{aric2008exploring} with minimum degree $k_{min}=5$, maximum degree ${k_{max} = 100}$ (scaled according to $k_{max}\simeq\sqrt{N}$ when changing $N$) and exponent $\gamma=2.1$. The resulting average degree is $\langle k \rangle=14.7$. Erdős–Rényi network are generated using NetworkX's implementation of \cite{batagelj2005efficient}, the probability of each link is set to $p=14.7/N$ to ensure the same average degree is obtained in both networks.

The equilibrium value of $a^j_t$ is determined by having individuals react ex-post to changes in $\textbf{a}_t \backslash a^j_t$ while keeping $\textbf{i}_t$ constant. More specifically, at each timestep, the health state of all individuals (equations \ref{eq: heterogeneous_model_s} - \ref{eq: heterogeneous_model_r} and \ref{eq: homogeneous_model_s} - \ref{eq: homogeneous_model_r} in the MF and HMF cases) is kept fixed, while social activities are updated by running Eq. \eqref{eq: social_activity_general} iteratively using $\textbf{a}_{n-1}\backslash a^j_{n-1}$ as an approximation for $\textbf{a}_{n} \backslash a^j_n$ ($n \in [1, N_{steps}]$) with $\textbf{a}_{0}\backslash a^j_0 = \textbf{a}_{t-1} \backslash a^j_{t-1}$. We use $N_{steps} = 10$ iterations and observe that, even for high values of $\alpha$, a fixed point ($\textbf{a}_{n} = \textbf{a}_{n-1}$) is typically reached after 4-5 iterations. The final value $\textbf{a}_{N_{steps}}$ is used to determine $\textbf{a}_{t}\backslash a^j_t$ in Eq. \eqref{eq: social_activity_general} (equations \ref{eq: homogeneous_model_a}, \ref{eq: heterogeneous_model_a}, \ref{eq: a_opt_quenched}).

We also tested the robustness of the model to changes in the size of the time step $dt$ observing no qualitative changes in the range $dt \in [0.1,1]$. The results presented in this paper are obtained for $dt=1$ and parameters are calibrated by assuming that $dt=1$ corresponds to a day as explained in Section \ref{sec: behavioral_parameters}. When changing $dt$, parameters are scaled according to $\mu=dt/10$, $\beta=0.3dt/\langle k \rangle$, $\delta=\exp{(-dt/10)}$ to ensure that the average duration of an infection, the basic reproduction number and the time constant of the exponential decay of the utility due to discounting remain unchanged. Moreover, to match the value of $a_{eq}$ corresponding to $dt=1$, we leave the average cost of infection $\alpha$ unchanged and scale the penalty for reducing the social activity (Eq. \ref{eq: U_a}) according to $U(a_t) = dt(\log(a_t) - a_t +1)$.

\end{document}